\documentclass[twocolumn,times]{aastex631}

\usepackage{amsmath}
\usepackage[T1]{fontenc}

\newcommand\ergs{erg~s$^{-1}$}
\newcommand\ergcms{erg~cm$^{-2}$~s$^{-1}$}
\newcommand\ergcmsa{erg~cm$^{-2}$~s$^{-1}$~arcsec$^{-2}$}
\newcommand\kms{km~s$^{-1}$}
\newcommand\cm{cm$^{-3}$}

\newcommand{\hbeta}{H{$\beta$}}
\newcommand{\halpha}{H{$\alpha$}}
\newcommand{\OIII}{[O\,{\sc iii}]}
\newcommand{\OIIIa}{[O\,{\sc iii}]\,$\lambda$4959}
\newcommand{\OIIIb}{[O\,{\sc iii}]\,$\lambda$5007}
\newcommand{\OIIIwave}{[O\,{\sc iii}]\,$\lambda\lambda$4959,5007}
\newcommand{\HeII}{He\,{\sc ii}}
\newcommand{\HeIIwave}{He\,{\sc ii}\,$\lambda$4686}
\newcommand{\OI}{[O\,{\sc i}]}

\newcommand{\NIIa}{[N\,{\sc ii}]\,$\lambda$6548}
\newcommand{\NIIb}{[N\,{\sc ii}]\,$\lambda$6583}
\newcommand{\SII}{[S\,{\sc ii}]}
\newcommand{\SIIwave}{[S\,{\sc ii}]\,$\lambda\lambda$6716,\,6731}
\newcommand{\SIIa}{[S\,{\sc ii}]\,$\lambda$6716}
\newcommand{\SIIb}{[S\,{\sc ii}]\,$\lambda$6731}

\shorttitle{Bubble nebulae around NGC 55 ULX-1}
\shortauthors{Zhou et al.}

\graphicspath{{./}{fig/}}

\begin{document}

\title{Identification of Bubble Nebulae around NGC 55 ULX-1 with MUSE Observations}

\author[0000-0002-5954-2571]{Changxing Zhou}
\affiliation{Department of Engineering Physics, Tsinghua University, Beijing 100084, China}

\author[0000-0001-7584-6236]{Hua Feng}
\affiliation{Department of Astronomy, Tsinghua University, Beijing 100084, China}

\author[0000-0002-1620-0897]{Fuyan Bian}
\affiliation{European Southern Observatory, Alonso de C\'{o}rdova 3107, Casilla 19001, Vitacura, Santiago 19, Chile}

\correspondingauthor{Hua Feng}
\email{hfeng@tsinghua.edu.cn}
\correspondingauthor{Fuyan Bian}
\email{fuyan.bian@eso.org}

\begin{abstract}
Using the Multi Unit Spectroscopic Explorer (MUSE) instrument on the Very Large Telescope, we identified three bubble nebulae (denoted as A, B, and C) around an ultraluminous X-ray source (ULX) in NGC 55. Bubble A shows a regular elliptical shape surrounding the ULX, with a morphology similar to the canonical ULX bubble around NGC 1313 X-2. It is most likely inflated by the ULX disk wind with a mechanical power close to $10^{39}$~\ergs. Bubble B lies 11\arcsec\ away from the ULX on the sky plane and is not contiguous to Bubble A. It displays a bow shock like morphology, and is likely driven by a collimated dark jet from the ULX with a mechanical power of about $3 \times 10^{38}$~\ergs. If this scenario is correct, we predict that Bubble B should present radio emission with a flux of about $1 - 10^2$~$\mu$Jy at 5~GHz. Bubble C appears within Bubble A, with a velocity and velocity dispersion distinct from the rest of Bubble A. Its nature is unclear and could be part of Bubble A as a result of low local density. The optical counterpart of ULX-1 exhibits a broad \halpha, consistent with emission from a hot disk wind.
\end{abstract}

\section{Introduction}

Ultraluminous X-ray sources (ULXs) are X-ray binaries with an apparent luminosity higher than $10^{39}$~\ergs. In general, they are thought to be powered by supercritical accretion onto neutron stars or stellar-mass black holes \citep[for reviews see][]{Kaaret2017,Fabrika2021,King2023,Pinto2023}. A remarkable feature for supercritical accretion is the presence of massive, ultrafast disk winds, which have been evidenced with X-ray spectroscopy \citep{Pinto2016,Pinto2017,Pinto2021,Kosec2018}, consistent with predictions from numerical simulations \citep[e.g.,][]{Jiang2014,Sadowski2014,Kitaki2021}. 

\begin{figure*}[tb]
\includegraphics[width=0.33\linewidth]{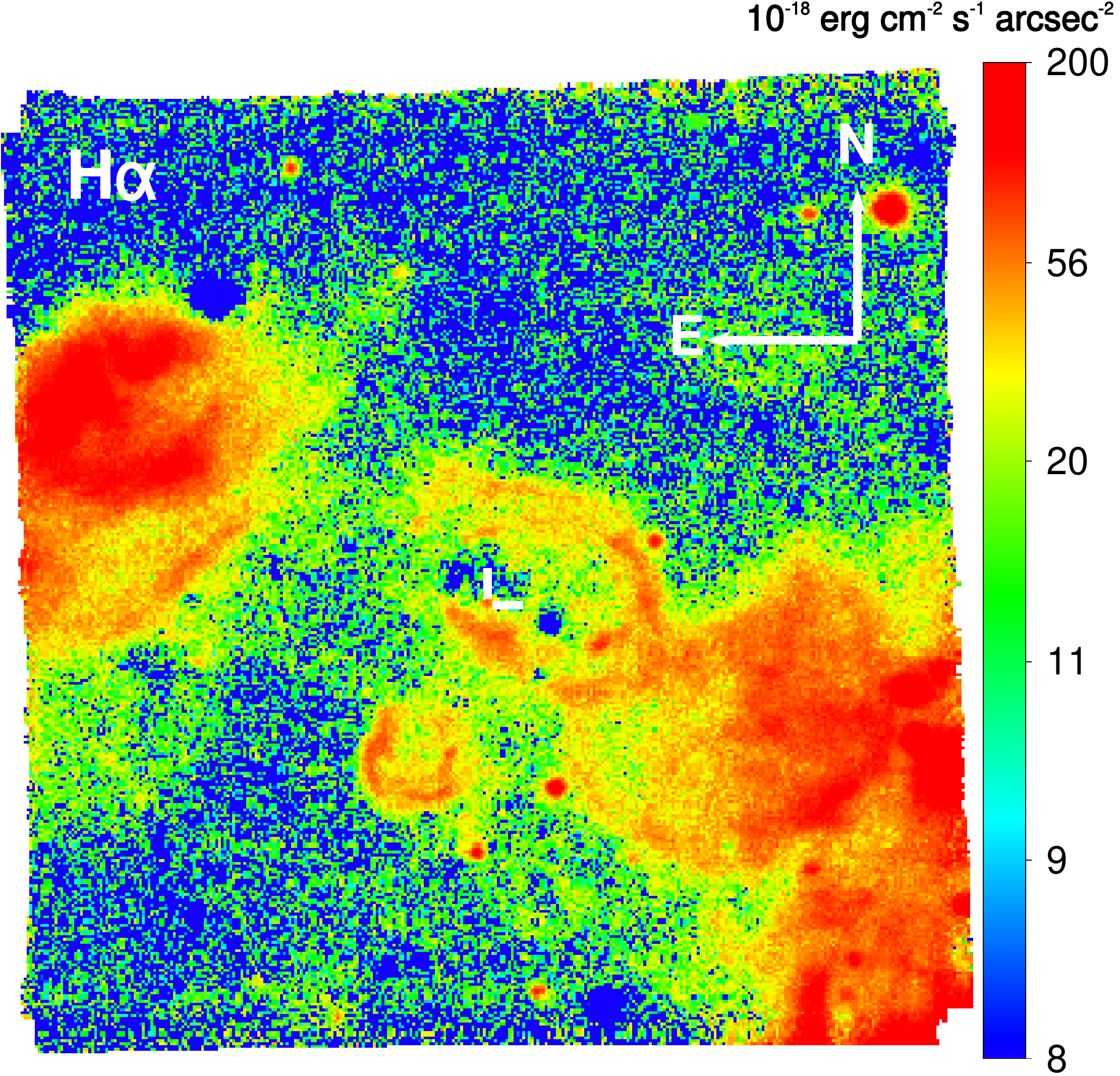}
\includegraphics[width=0.33\linewidth]{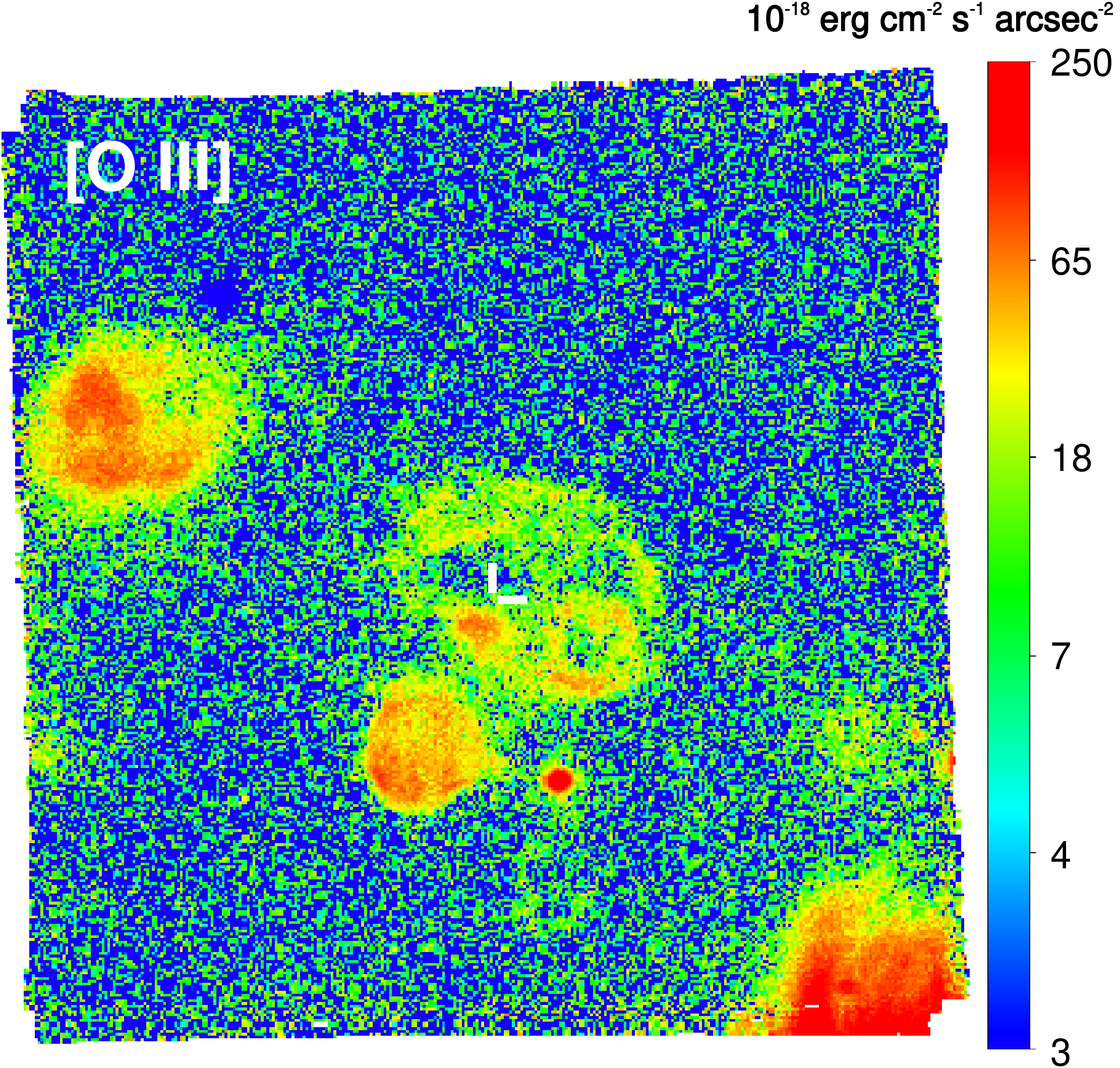}
\includegraphics[width=0.33\linewidth]{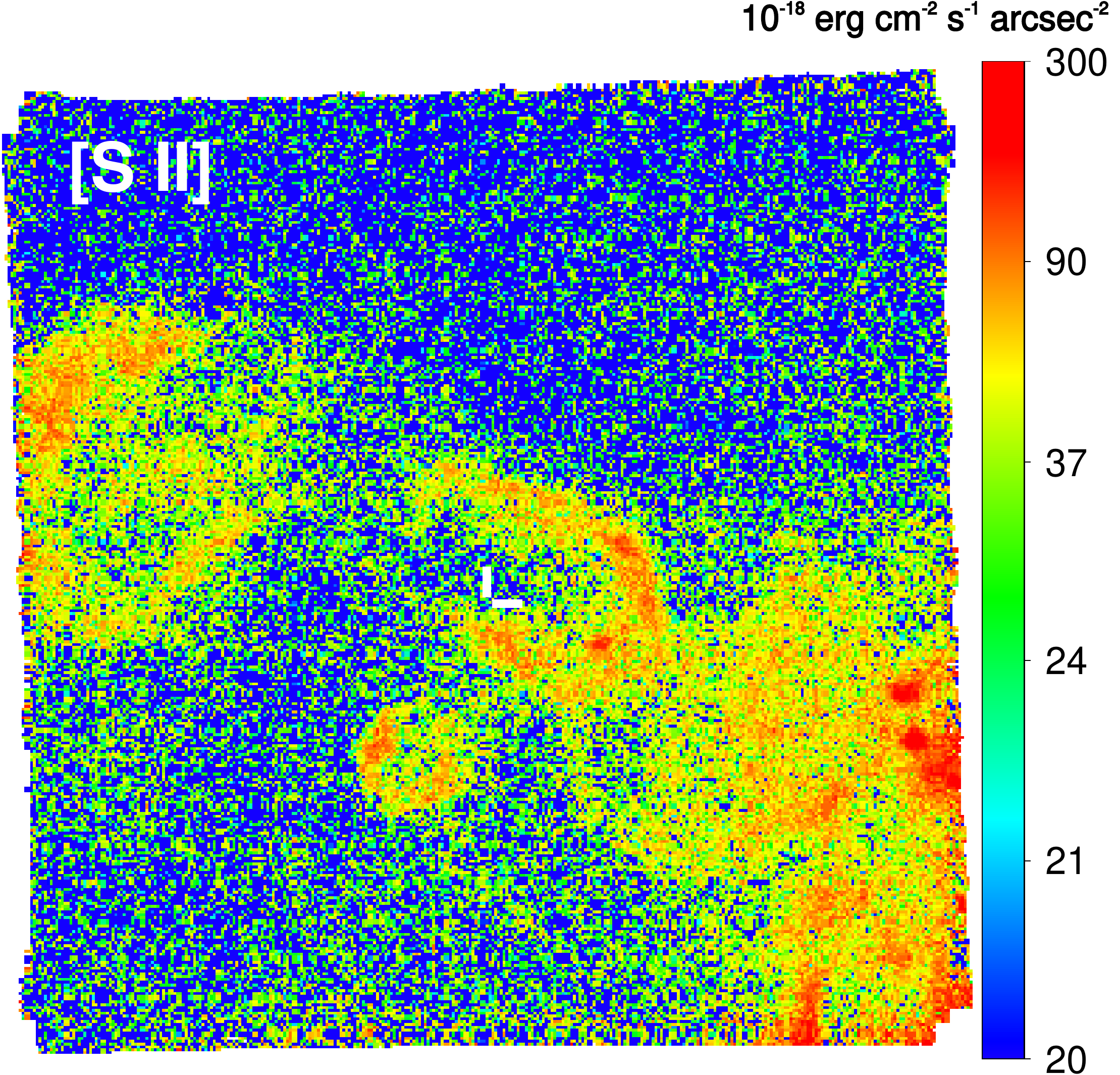}
\includegraphics[width=0.33\linewidth]{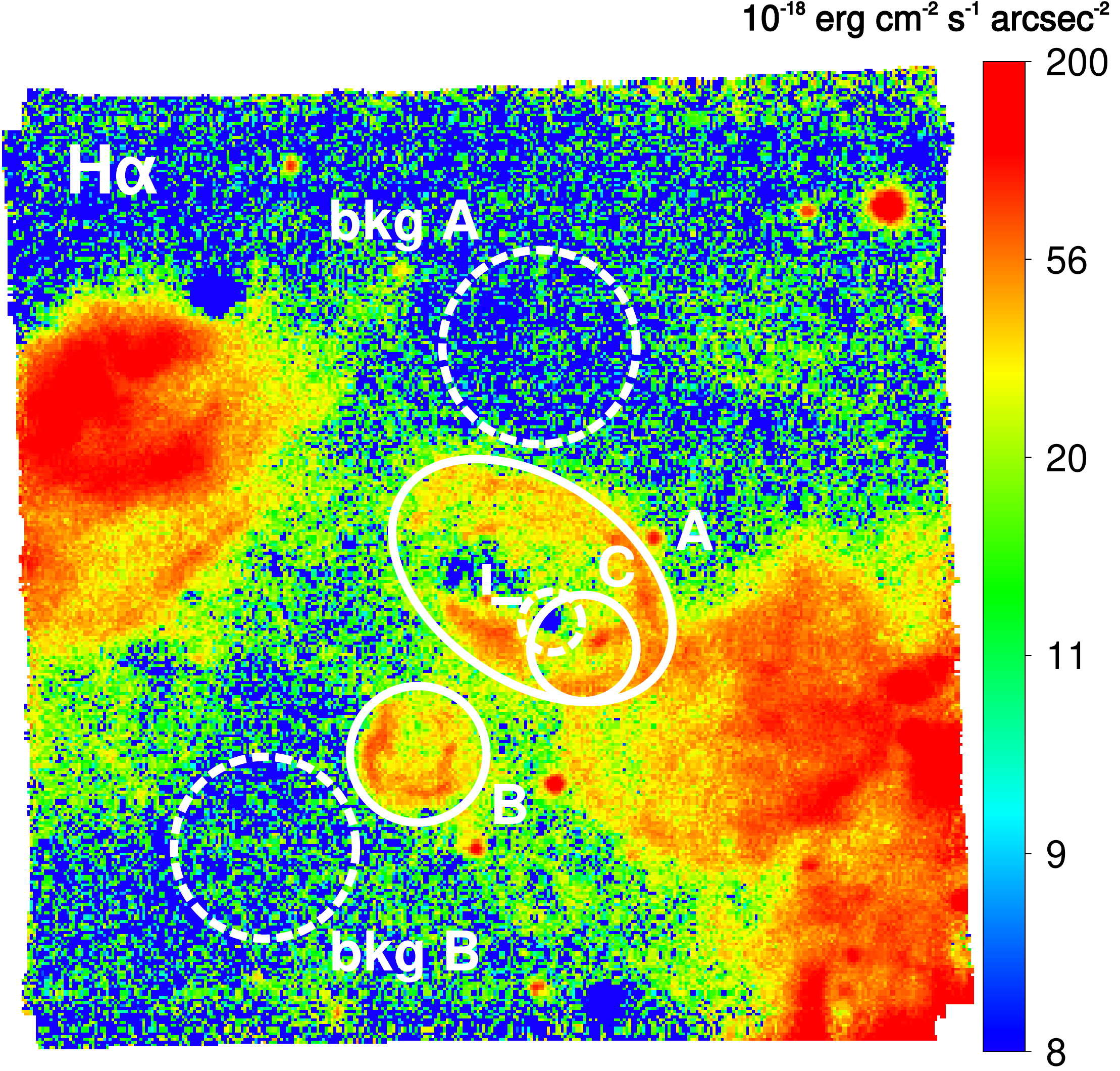}
\includegraphics[width=0.33\linewidth]{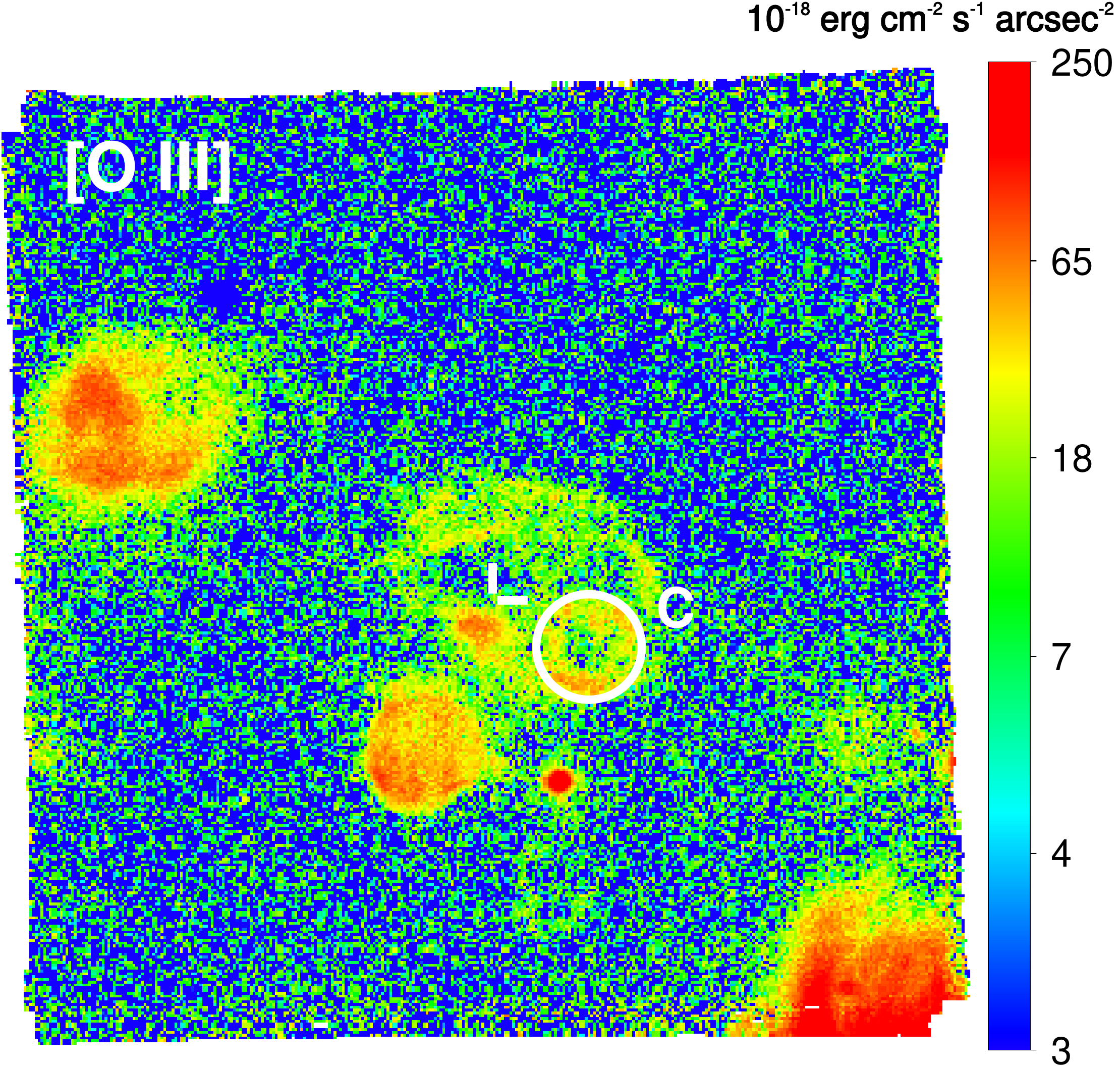}
\includegraphics[width=0.33\linewidth]{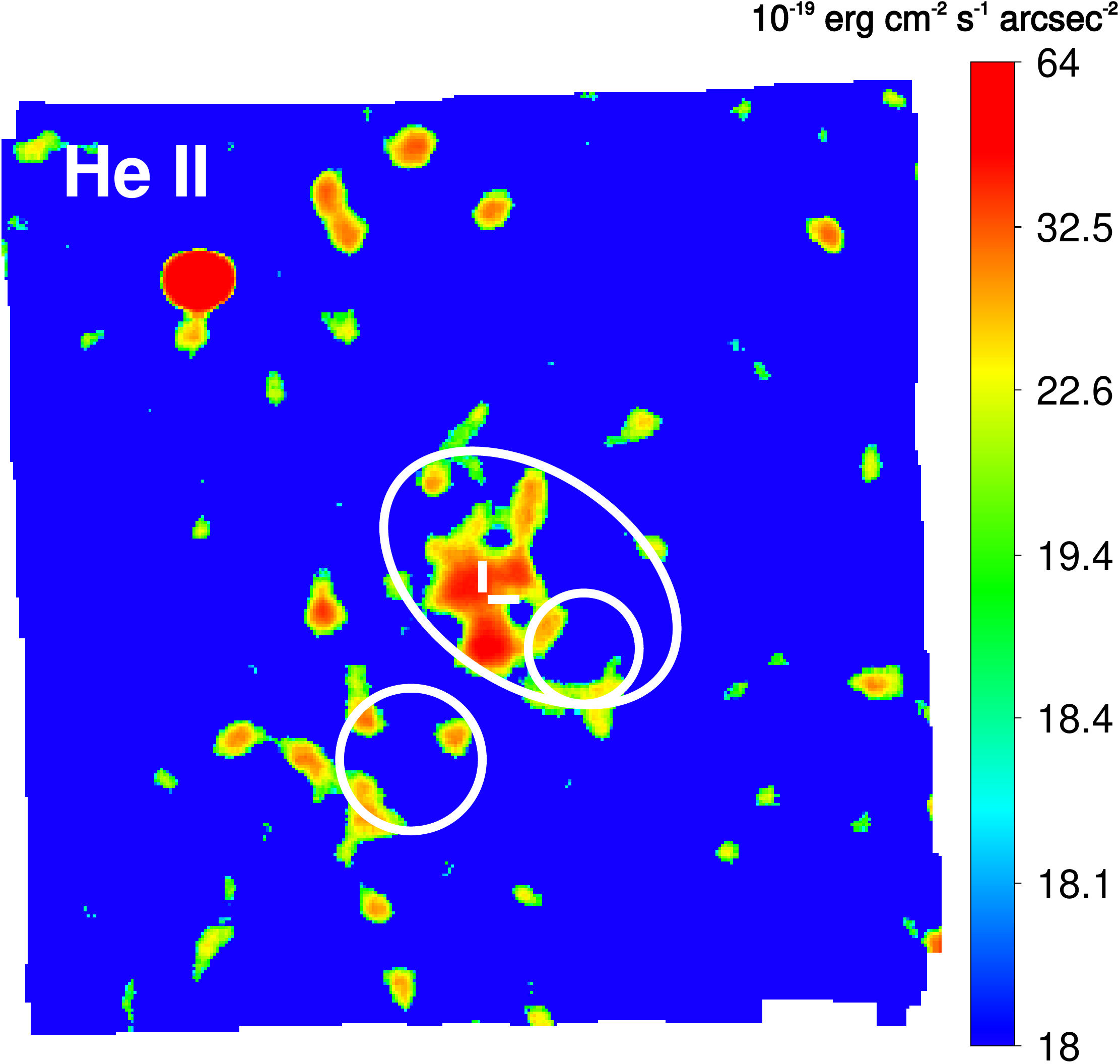}
\caption{MUSE emission flux images around NGC 55 ULX-1 for \halpha, \OIIIb, \SIIa, and \HeIIwave. The line flux is extracted from a 10~\AA\ window around the line centroid, after subtraction of continuum estimated from nearby wavelengths. A Gaussian smoothing with $\sigma = 5$~pixels or 1\arcsec\ is applied in the \HeII\ image. The bars mark the ULX-1 position. The three bubble nebulae are marked as A, B, and C, respectively. The two ``bkg'' regions are used for background estimate and subtraction, respectively for Bubble A/C and Bubble B. The dashed circle within Bubble A is to encircle a star cluster where the data are discarded. The arrows points north and east and have a length of 10\arcsec.}
\label{fig:img}
\end{figure*}

Optical observations of ULXs may help place constraints on their compact object mass \citep{Liu2013,Motch2014}, accretion geometry \citep{Vinokurov2013,Ambrosi2018,Yao2019}, history of binary evolution \citep{Madhusudhan2008,Patruno2008,Zhou2023}, and interaction with the environment \citep[e.g.,][]{Pakull2002}, and are thus an important probe to their physical nature. ULXs show optical emission lines reminiscent of luminous blue variables or late nitrogen Wolf-Rayet stars, or the supercritically accreting microquasar SS~433, which are all objects with powerful winds, suggesting that the optical emission lines seen in ULXs originate from a hot disk wind \citep{Fabrika2015}. 

If these winds with a high mechanical power are injected into the interstellar medium, shocks will be produced appearing as an optical nebula \citep{Siwek2017}. Indeed, optical bubble nebulae with a size of some 100~pc have been observed around many ULXs \citep[e.g.,][]{Pakull2002,Pakull2003,Pakull2006,Pakull2008,Ramsey2006,Abolmasov2007,Abolmasov2008,Russell2011,Cseh2012,Urquhart2018,Soria2021,Gurpide2022,Zhou2022,Guo2023}. They exhibit bright \SII\ and \OI\ emission, typical for shock ionization. Based on the \OIIIb/\hbeta\ and other line flux ratios, a shock velocity of about 100~\kms\ can be inferred, indicative of supersonic expansion. The emission line luminosity, bubble size, and shock velocity suggest that the wind or jet that drives the bubble has a mechanical power in the range of $\sim$$10^{39}$--$10^{40}$~\ergs\ and an age typically of Myr. Thus, ULXs are ultraluminous in terms of both radiation and mechanical output. In some cases, photoionization is a non-negligible addition to shock ionization \citep[e.g.,][]{Abolmasov2007,Urquhart2018,Gurpide2022,Zhou2022} and even the dominant source in powering the emission line nebula \citep[e.g.,][]{Pakull2002,Kaaret2004,Pakull2008,Kaaret2009}. Bubbles with a super-Eddington mechanical power are also found around non-ULXs \citep[e.g.,][]{Pakull2010,Soria2014}, similar to the W50 nebula around the Galactic microquasar SS~433 \citep{Fabrika2004}. 

NGC 55 is an edge-on Magellanic type barred spiral galaxy located in the Sculptor group. ULX-1 is the brightest persistent X-ray source in it with a luminosity typically around a few times $10^{39}$~\ergs\ \citep{Stobbart2004,Binder2015,Pintore2015,Pinto2017,Gurpide2021,Jithesh2022,Barra2022}. NGC 55 ULX-1 displays a soft X-ray spectrum and makes transitions between the soft and supersoft regimes \citep{Pinto2017,Gurpide2021,Barra2022}. Its X-ray lightcurve occasionally exhibits sharp drops and dips \citep{Stobbart2004,Barra2022}. High-resolution X-ray spectroscopy revealed multiple components of emission and absorption lines at different velocities, indicative of disk winds in a complex dynamical structure \citep{Pinto2017}. These X-ray behaviors argue in favor of the scenario that NGC 55 ULX-1 is viewed at a relatively large inclination angle similar to other soft or supersoft ULXs \citep{Feng2016,Urquhart2016a}. 

Despite extensive X-ray studies, optical observations of NGC 55 ULX-1 have been insufficient \citep[e.g.][]{Gladstone2013}. Here in this work, we present optical integral field spectroscopy of NGC 55 ULX-1 and its neighborhood with the Multi Unit Spectroscopic Explorer (MUSE) instrument \citep{Bacon2010} on the Very Large Telescope (VLT). We discovered bubble nebulae surrounding and near the ULX. The observations, data analysis and results are described in Section~\ref{sec:data}, and the possible physical nature is discussed in Section~\ref{sec:discuss}. To be in line with previous studies, we adopt a distance of 1.78~Mpc to NGC 55 \citep{Karachentsev2003}. 

\section{Observations and results}
\label{sec:data} 

NGC 55 ULX-1 was observed with VLT MUSE on the night of May 25th, 2022, at the La Silla Paranal Observatory under the program 109.238W.002. The observation follows the object-sky-object-object-sky-object pattern and contains $4 \times 560$~s on-source exposures and $2 \times 120$~s sky exposures. A 90-degree rotation plus small offset dithering patterns were applied between the on-source exposures. The seeing was about 0\farcs5 during the observations. The sky transparency conditions were partially cloudy, which had no impact to our target. The instrument mode used was WFM-NOAO-E, covering a wavelength range of 4600--9350~\AA.

The data were reduced by the EsoRex pipeline \citep{Weilbacher2020}. For each independent exposure, we used the {\it muse\_scibasic} recipe to carry out the standard correction of the bias frames, lamp flats, arc lamps, twilight flats, geometry, and illumination exposures. With standard stars taken on the same night, {\it muse\_scipost} was employed to correct the telluric and flux and subtract the sky using the separate sky exposure. We used {\it muse\_exp\_align} to align each exposure, and {\it muse\_exp\_combine} to generate the final data cube. The field of view of the final data cube is $1\arcmin \times 1\arcmin$, with the spaxel scale $0\farcs2 \times 0\farcs2$.

\begin{figure*}[bt]
\centering
\includegraphics[width=0.33\linewidth]{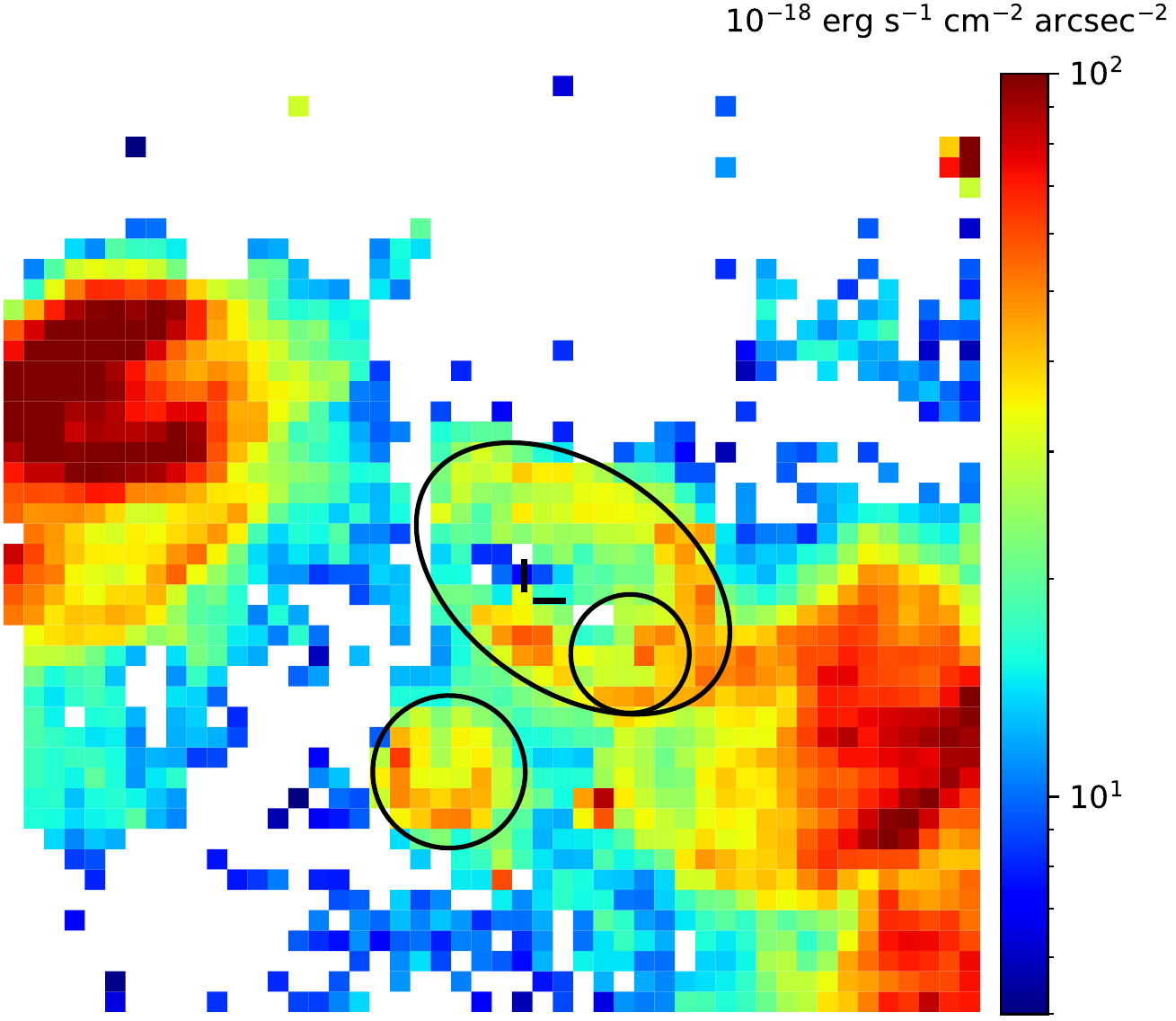}
\includegraphics[width=0.33\linewidth]{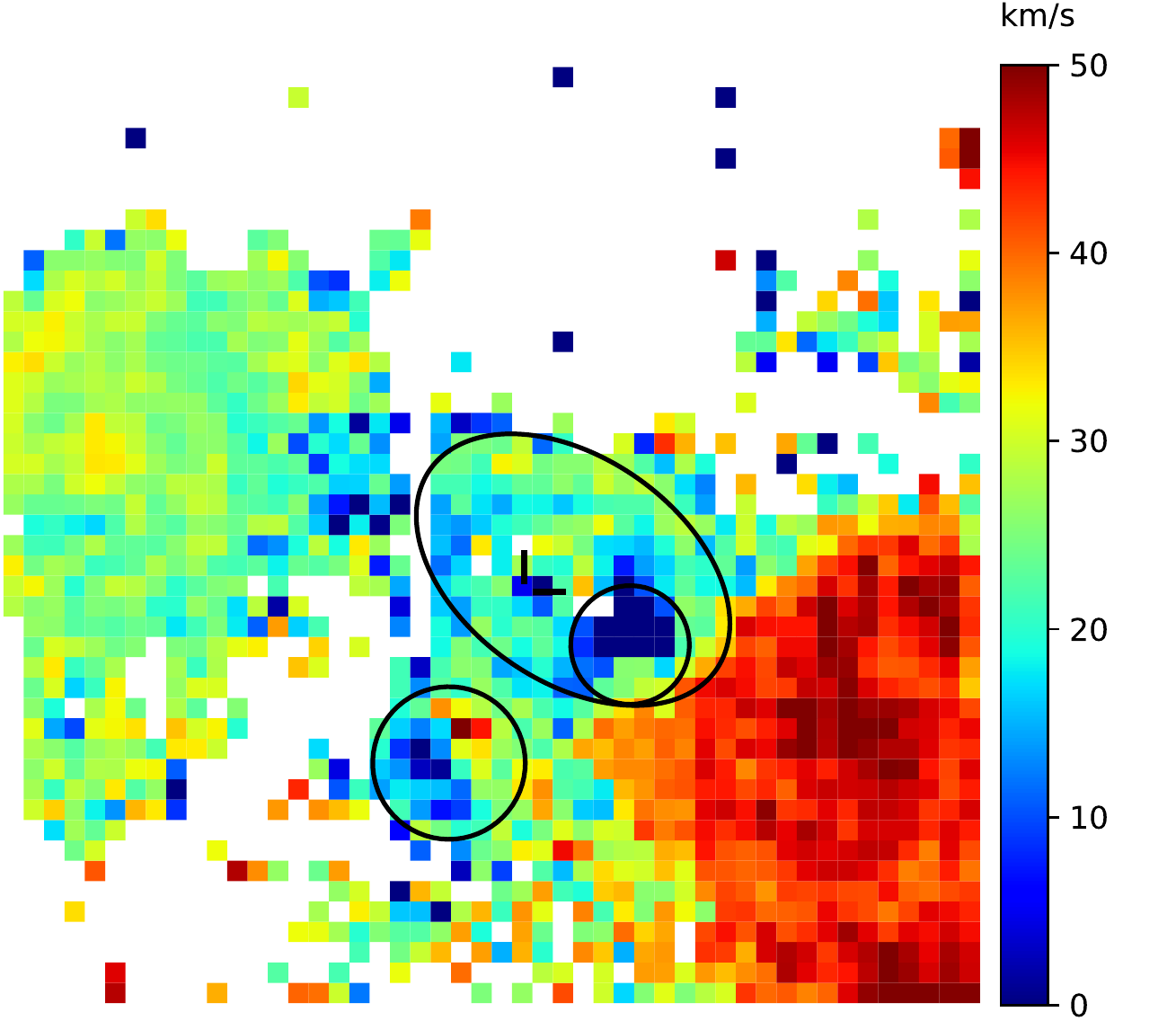}
\includegraphics[width=0.33\linewidth]{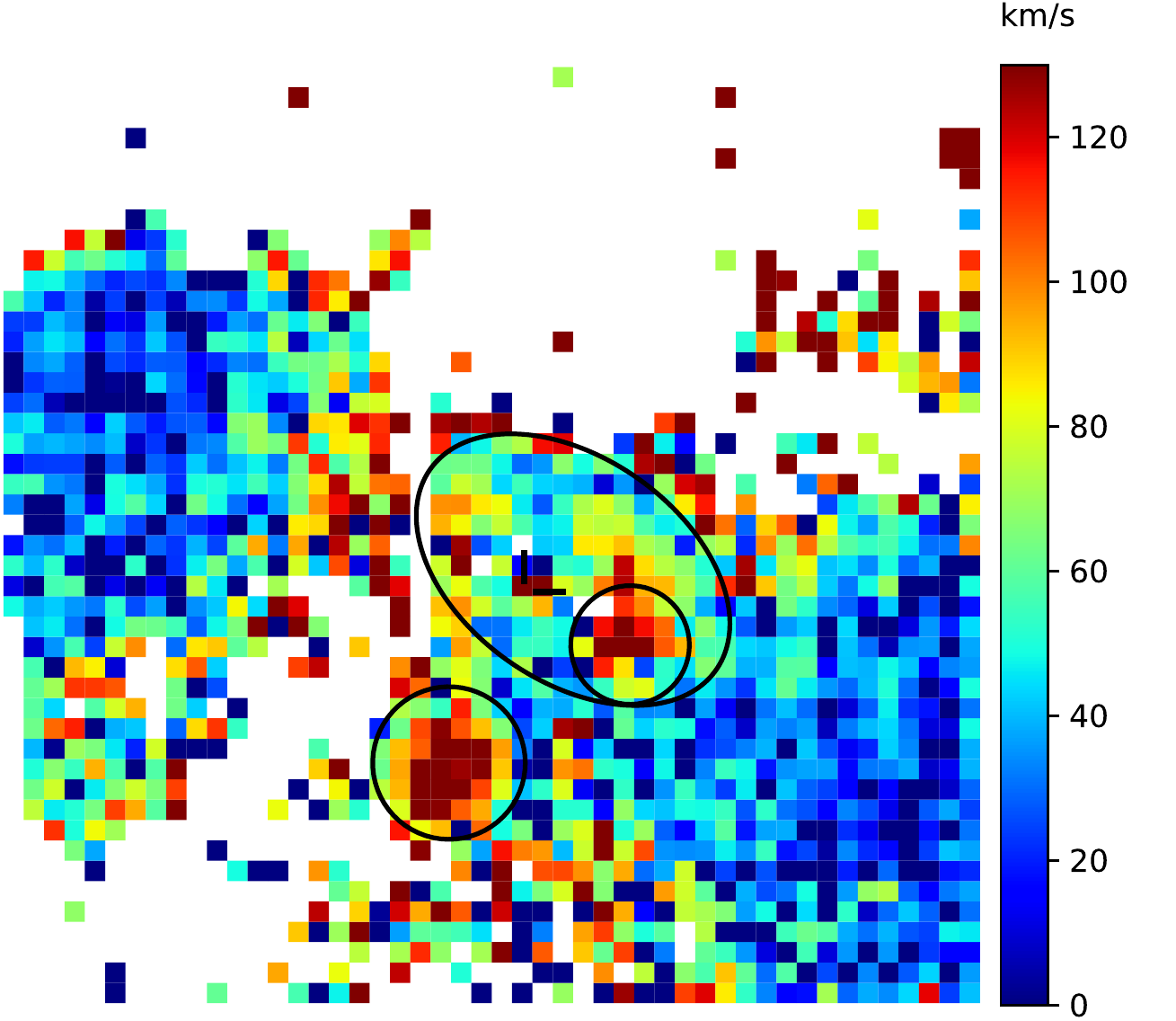}
\caption{\halpha\ line flux ({\bf left}), velocity with respect to the motion of the host galaxy ({\bf middle}), and FWHM line dispersion ({\bf right}). A binning of $6 \times 6$ pixels is applied to improve the S/N. Binned pixels with S/N $<$ 2 are not shown. The three bubbles and ULX are indicated.}
\label{fig:map_halpha}
\end{figure*}

\begin{figure*}
\centering
\includegraphics[width=0.33\linewidth]{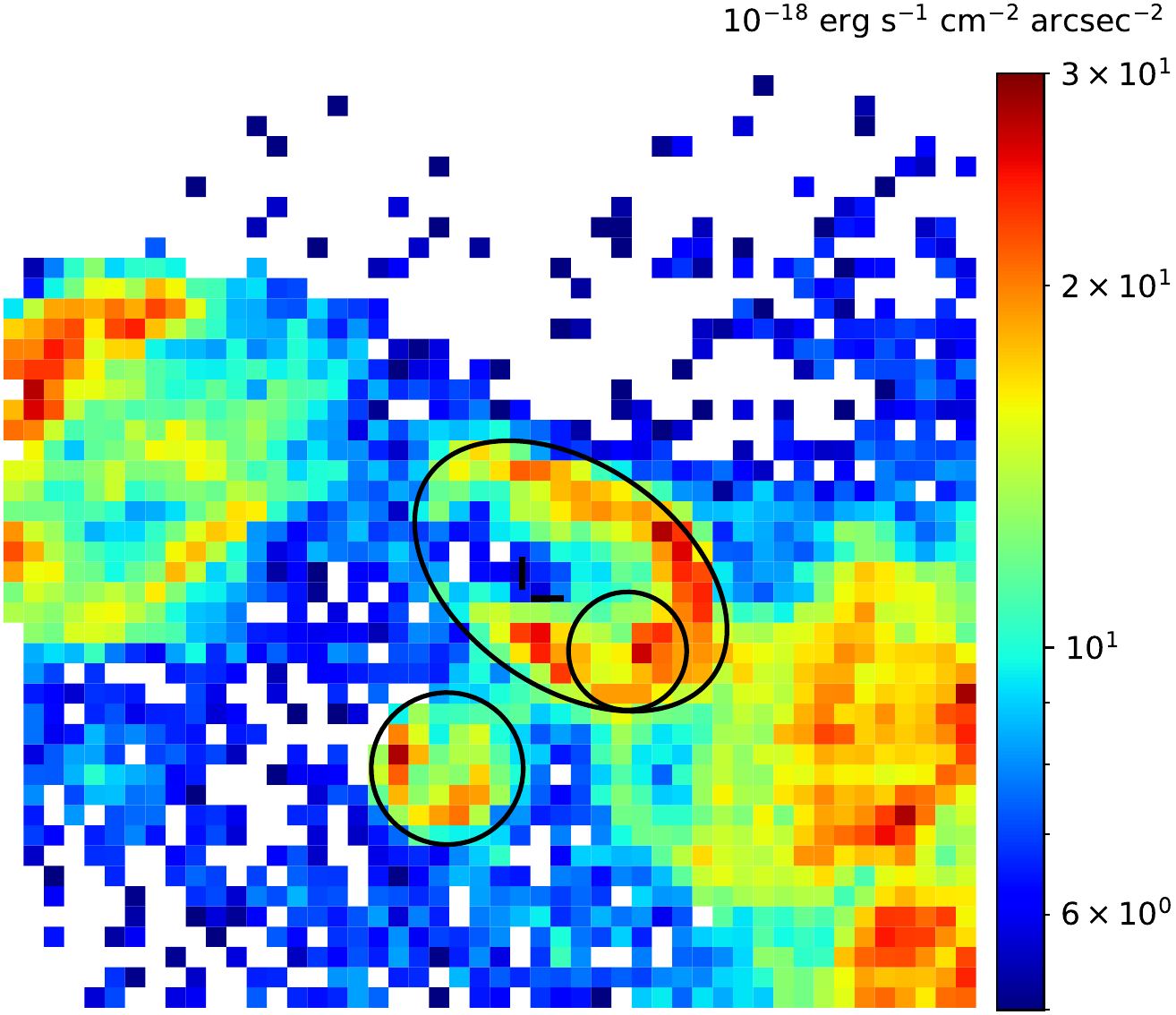}
\includegraphics[width=0.33\linewidth]{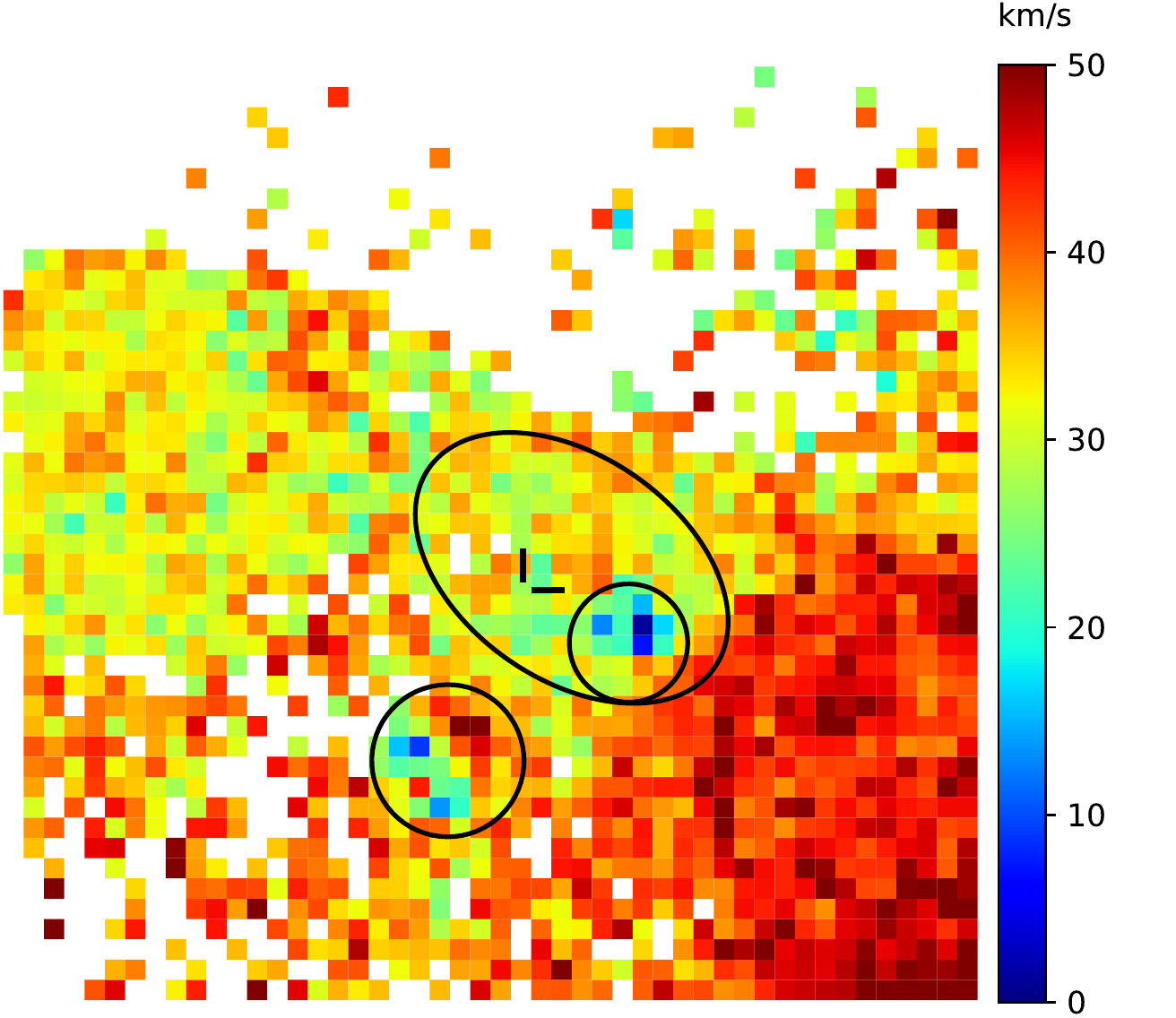}
\includegraphics[width=0.33\linewidth]{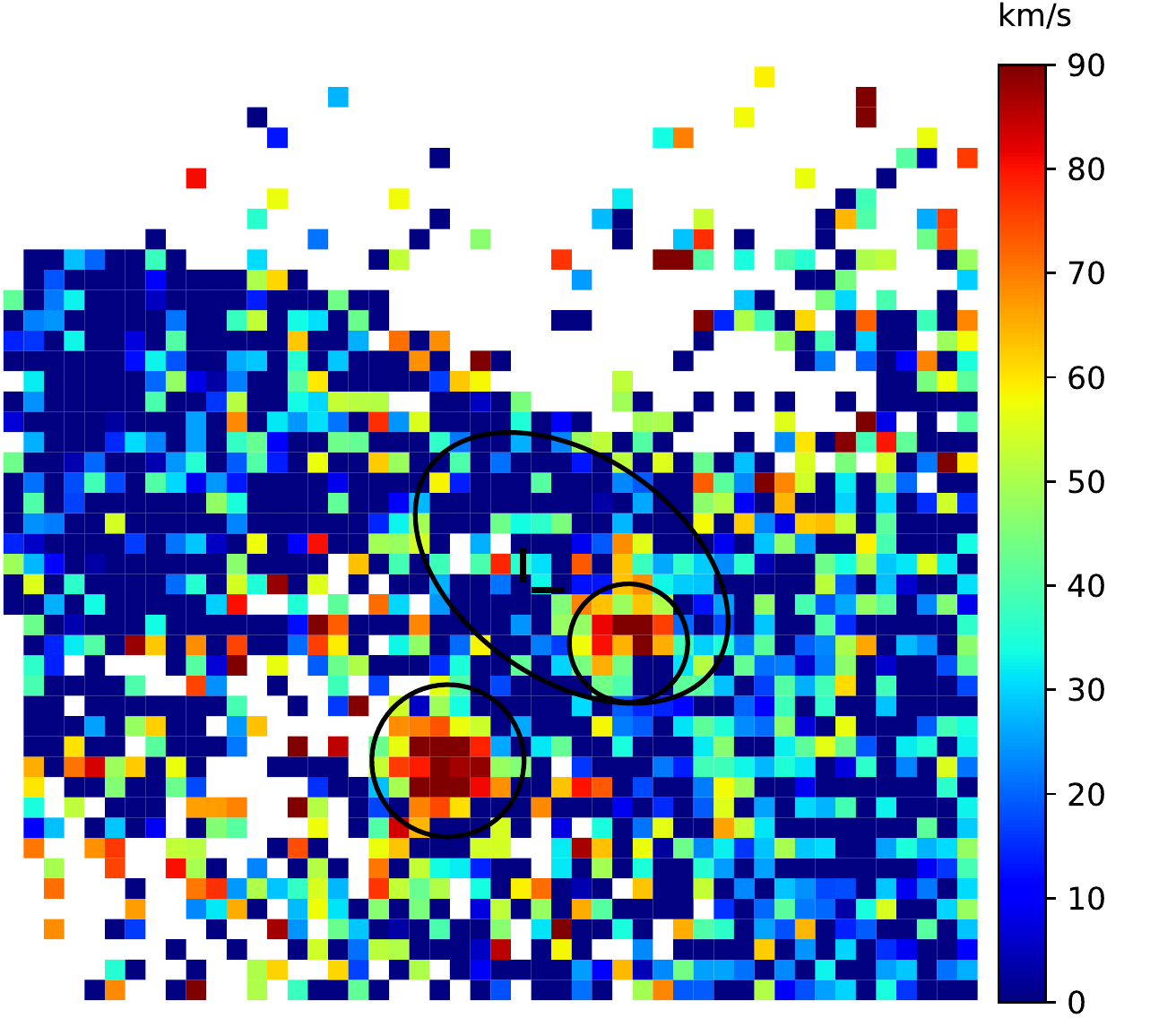}
\caption{Same as Figure~\ref{fig:map_halpha} but for \SII.}
\label{fig:map_sii}
\end{figure*}

\begin{figure*}
\centering
\includegraphics[width=0.33\linewidth]{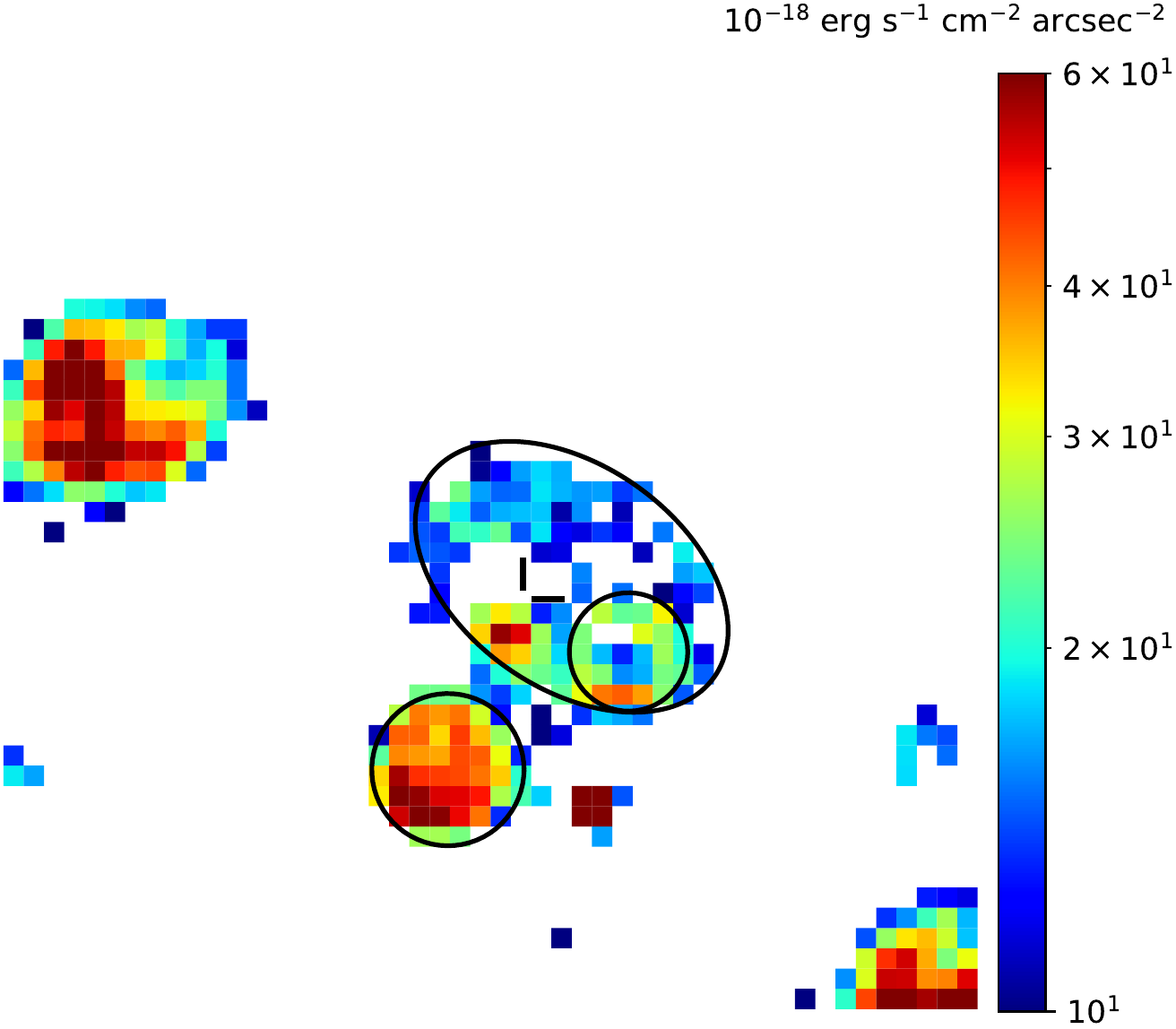}
\includegraphics[width=0.33\linewidth]{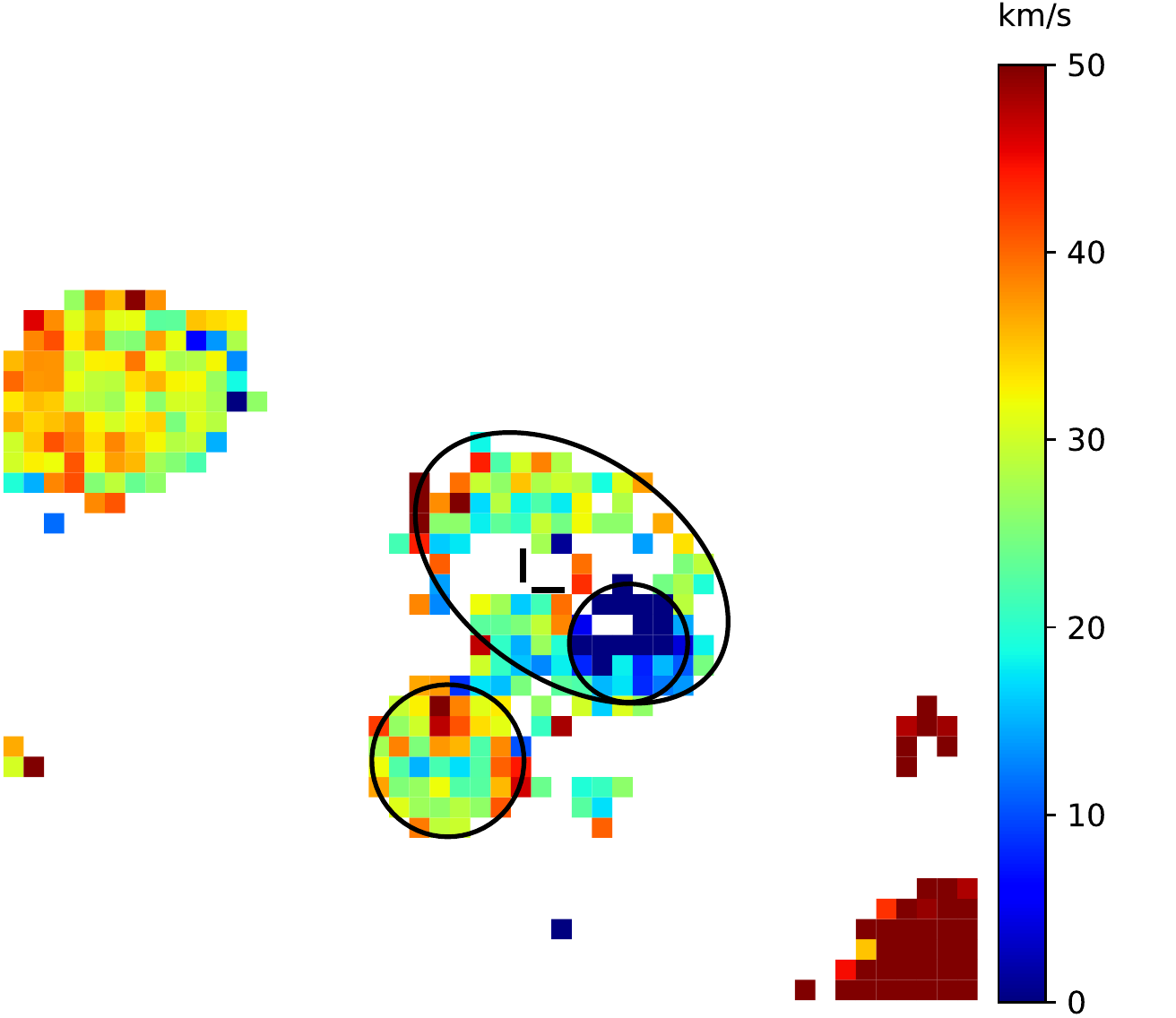}
\includegraphics[width=0.33\linewidth]{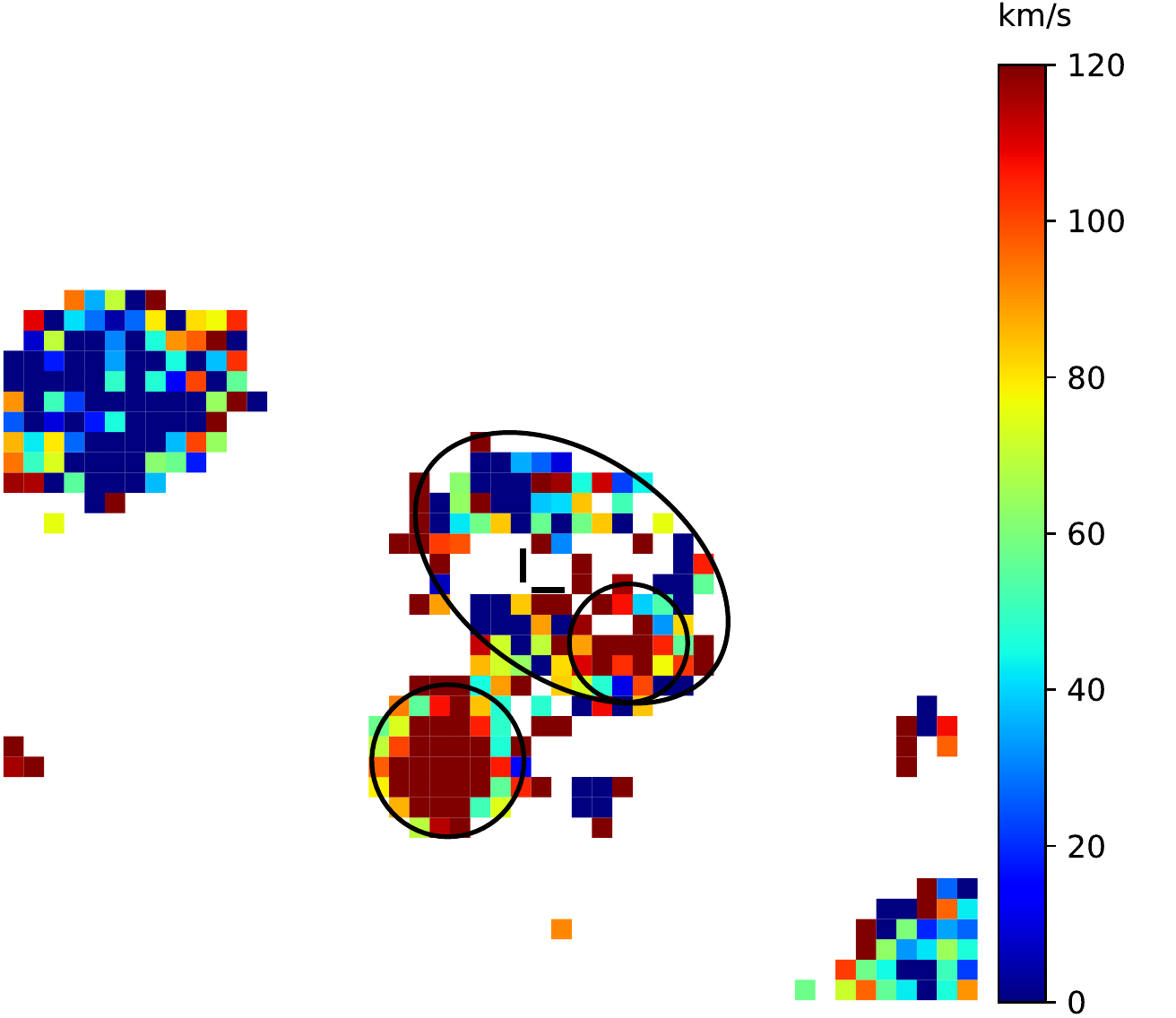}
\caption{Same as Figure~\ref{fig:map_halpha} but for \OIII.}
\label{fig:map_oiii}
\end{figure*}

Figure~\ref{fig:img} shows the \halpha, \OIIIb, \SIIa, and \HeIIwave\ images with fluxes directly extracted from the spectrum on each pixel. Three bubble nebulae are revealed: (A) an elliptical bubble with an extension of $21\arcsec \times 13\arcsec$ surrounding the ULX, (B) a nearly circular bubble with a diameter of 9\arcsec\ to the southeast of the ULX, and (C) a possible circular bubble within Bubble A with a diameter of 7\arcsec\ best visible in \OIII. They are indicated in Figure~\ref{fig:img}. Specifically, Bubbles A and B can be outlined with an \halpha\ brightness higher than $2 \times 10^{-17}$~\ergcmsa, and Bubble C isolates with \OIIIb\ higher than $3 \times 10^{-17}$~\ergcmsa. We note that there is a possible star cluster with strong continuum emission and Balmer absorption features in NGC 55 along the line of sight to Bubble A \citep[see Figure~1 in][]{Gladstone2013}. It shows a radial velocity consistent with that of the southwestern star forming region, indicating that it is not physically associated with the bubbles. It is encircled with a dashed region in Figure~\ref{fig:img} (within Bubble A and overlapped with Bubble C), and the data in it are discarded in this work. To the west of Bubble B and south of Bubble C, there is a bright, spatially unresolved object shown in \halpha\ and \OIII\ images. It has no obvious counterparts in the Hubble Space Telescope (HST) images, and has narrow (unresolved) line width. It shows a velocity significantly different from that of Bubble B. The luminosities and flux ratios of the emission lines indicate that it may be a planetary nebula in NGC 55 \citep{Frew2010,Parker2022,Ritter2023}.

For \halpha, the \SII\ doublet, and the \OIII\ doublet, which have the highest signal-to-noise (S/N) ratios, we extracted the line properties over a $6 \times 6$ pixel binning. Each line is fitted with a Gaussian. The continuum is first determined from  nearby wavelength intervals without emission lines and then subtracted. The doublet are fitted jointly forcing the same velocity and width. The flux ratio is fixed at 2.98 for \OIIIwave\ \citep{Storey2000}, but set free for \SIIwave. The maps for line flux, velocity, and velocity dispersion (always in FWHM and with instrument broadening corrected in this work) are shown in Figure~\ref{fig:map_halpha}, \ref{fig:map_sii}, and \ref{fig:map_oiii}, respectively, for \halpha, \SII, and \OIII. In this work, the velocity refers to the relative radial motion with respect to the host galaxy ($z = 0.00043$), and a negative velocity means blueshift. On the velocity and velocity dispersion maps, the three bubbles are distinct from the western star forming region, with more blueshift and a higher velocity dispersion, suggesting that they are not physically associated. Bubble C is also distinct from Bubble A in the same way, perhaps justifying an individual identification. It is also possible that Bubble C is a substructure of Bubble A. In the following, we provide the analysis results for both Bubble A (including C) and Bubble A $-$ C (without C).

\begin{figure}
\centering
\includegraphics[width=\columnwidth]{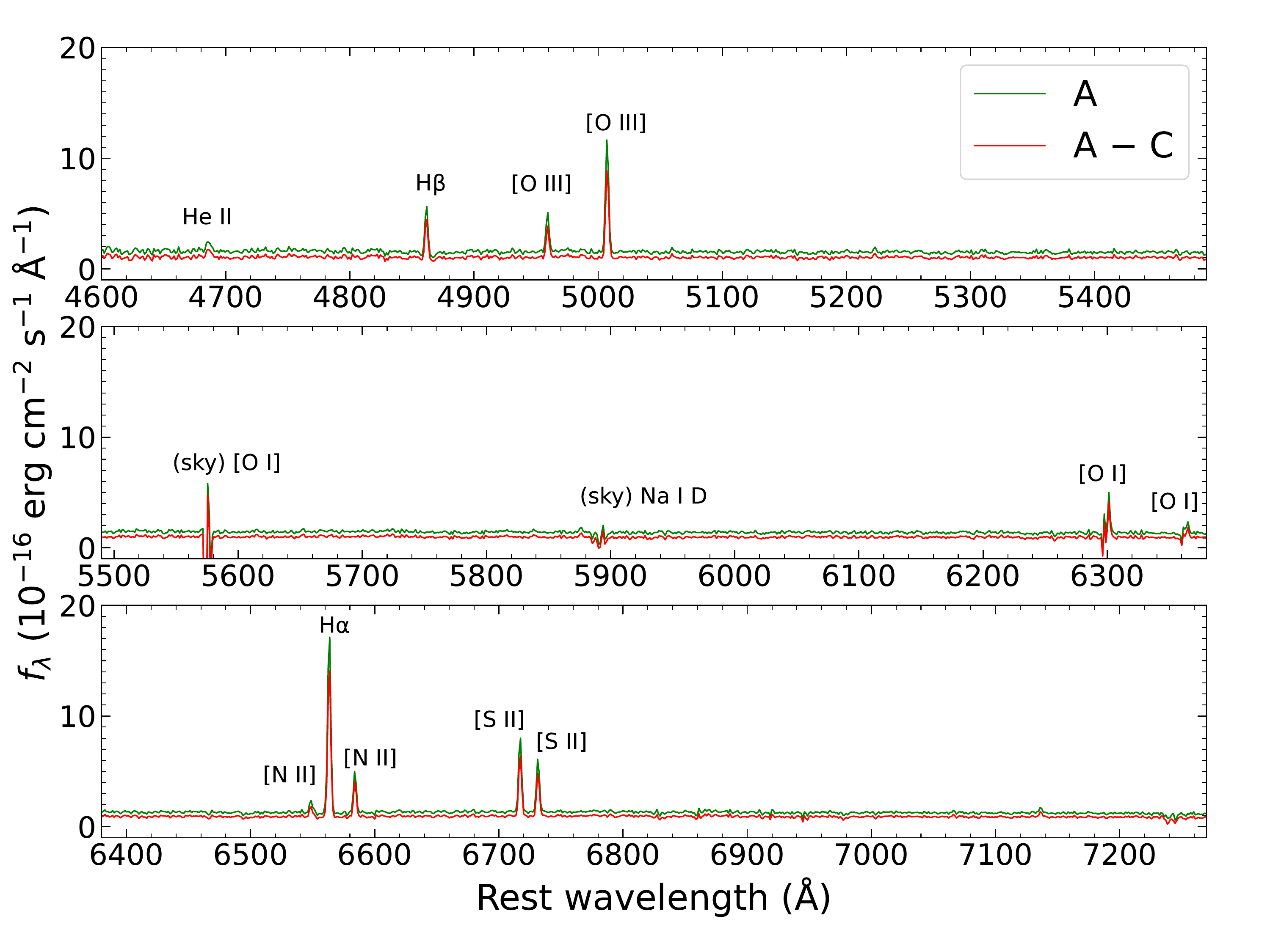}
\includegraphics[width=\columnwidth]{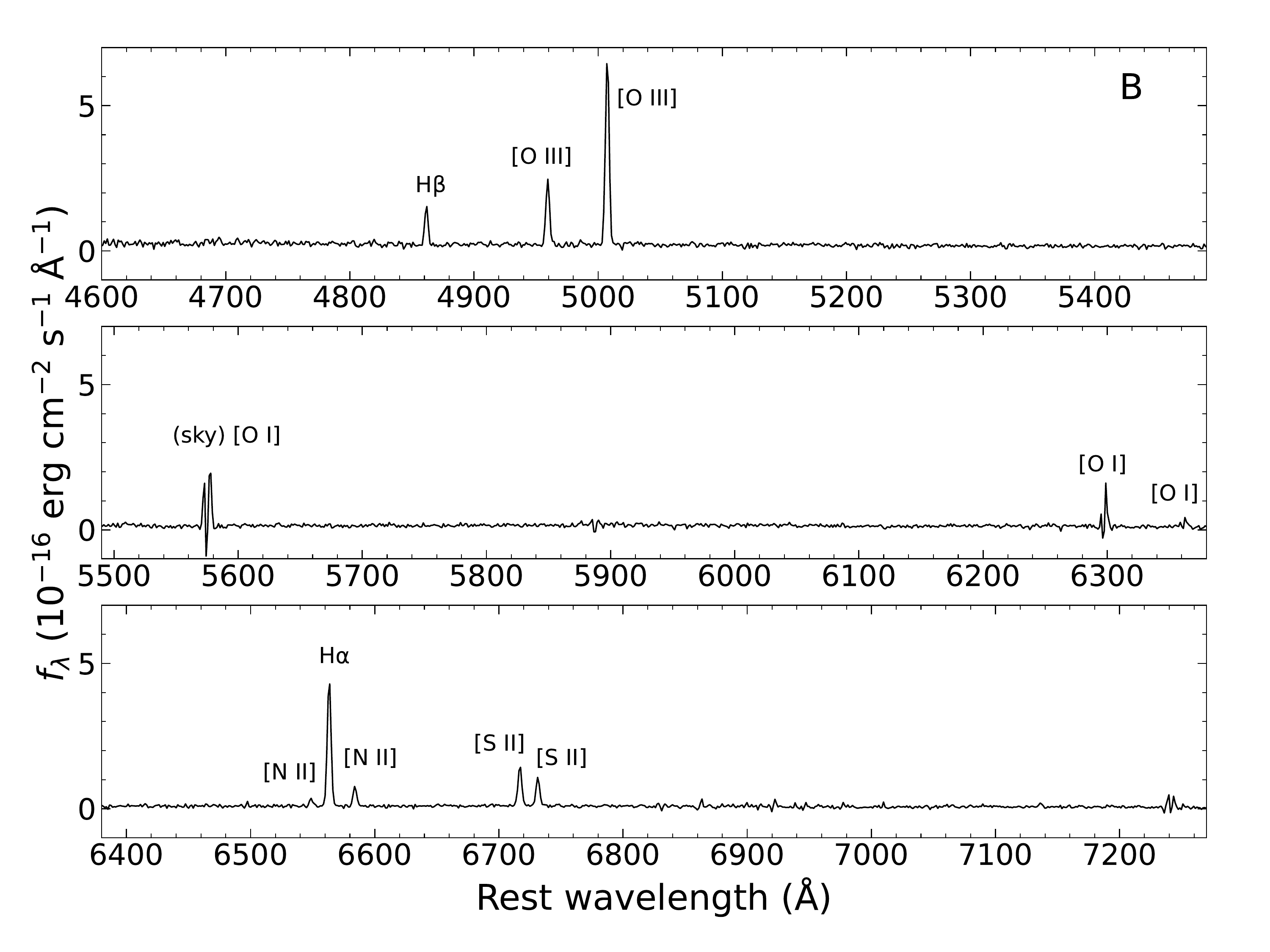}
\includegraphics[width=\columnwidth]{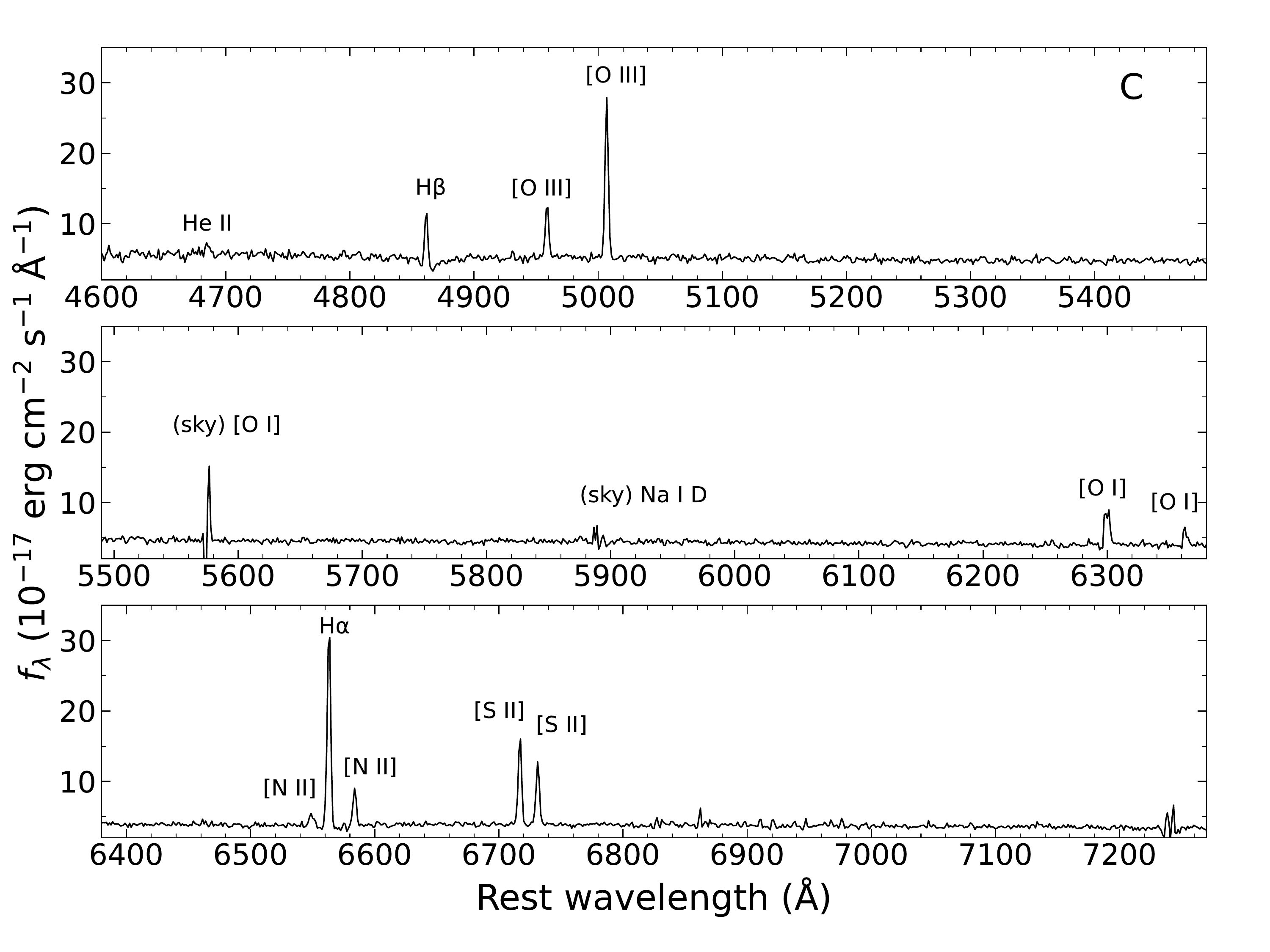}
\caption{Optical spectra for Bubbles A, A $-$ C, B, and C.}
\label{fig:spec_bubbles}
\end{figure}

\begin{figure}
\centering
\includegraphics[width=\columnwidth]{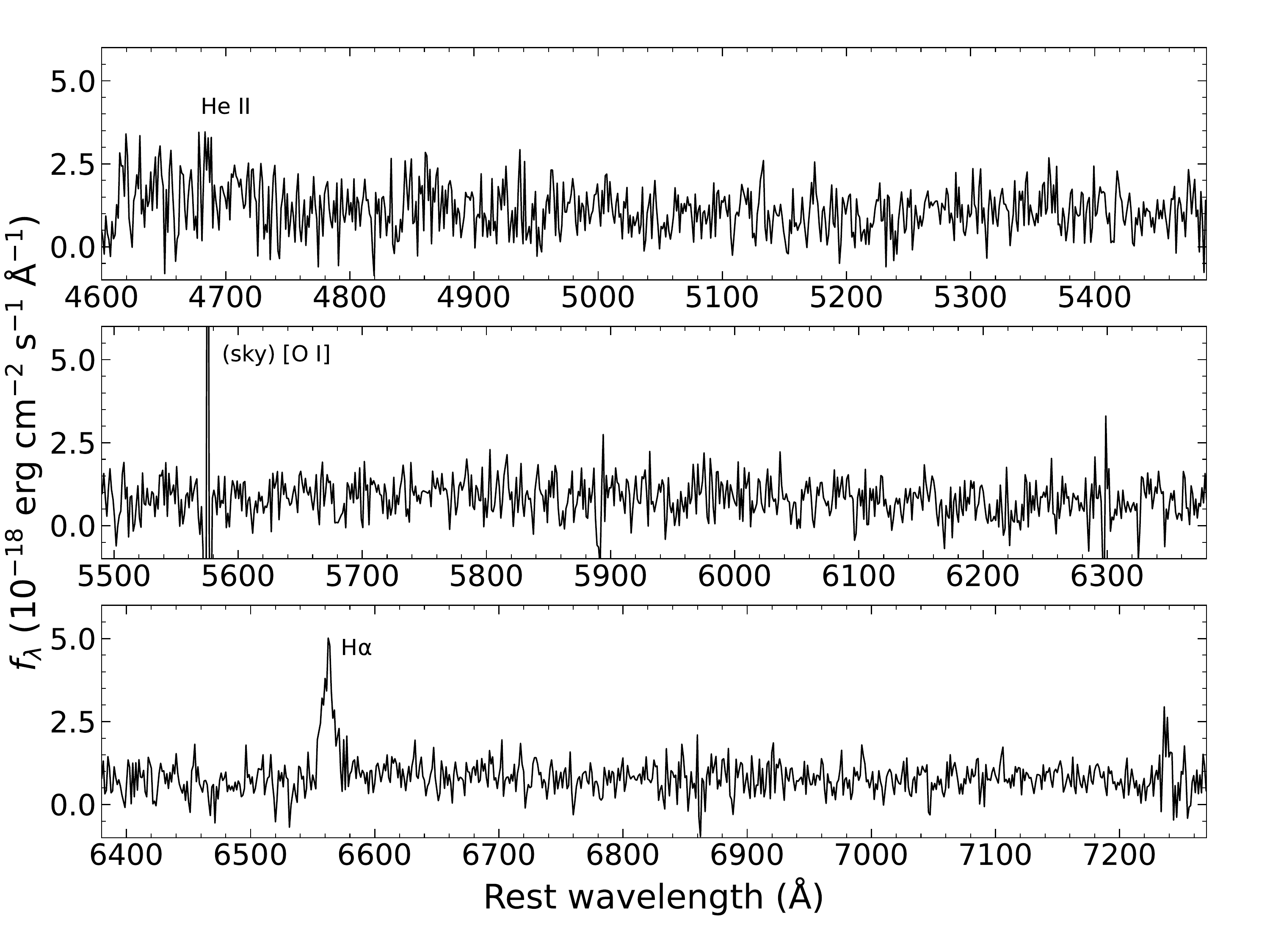}
\includegraphics[width=0.5\columnwidth]{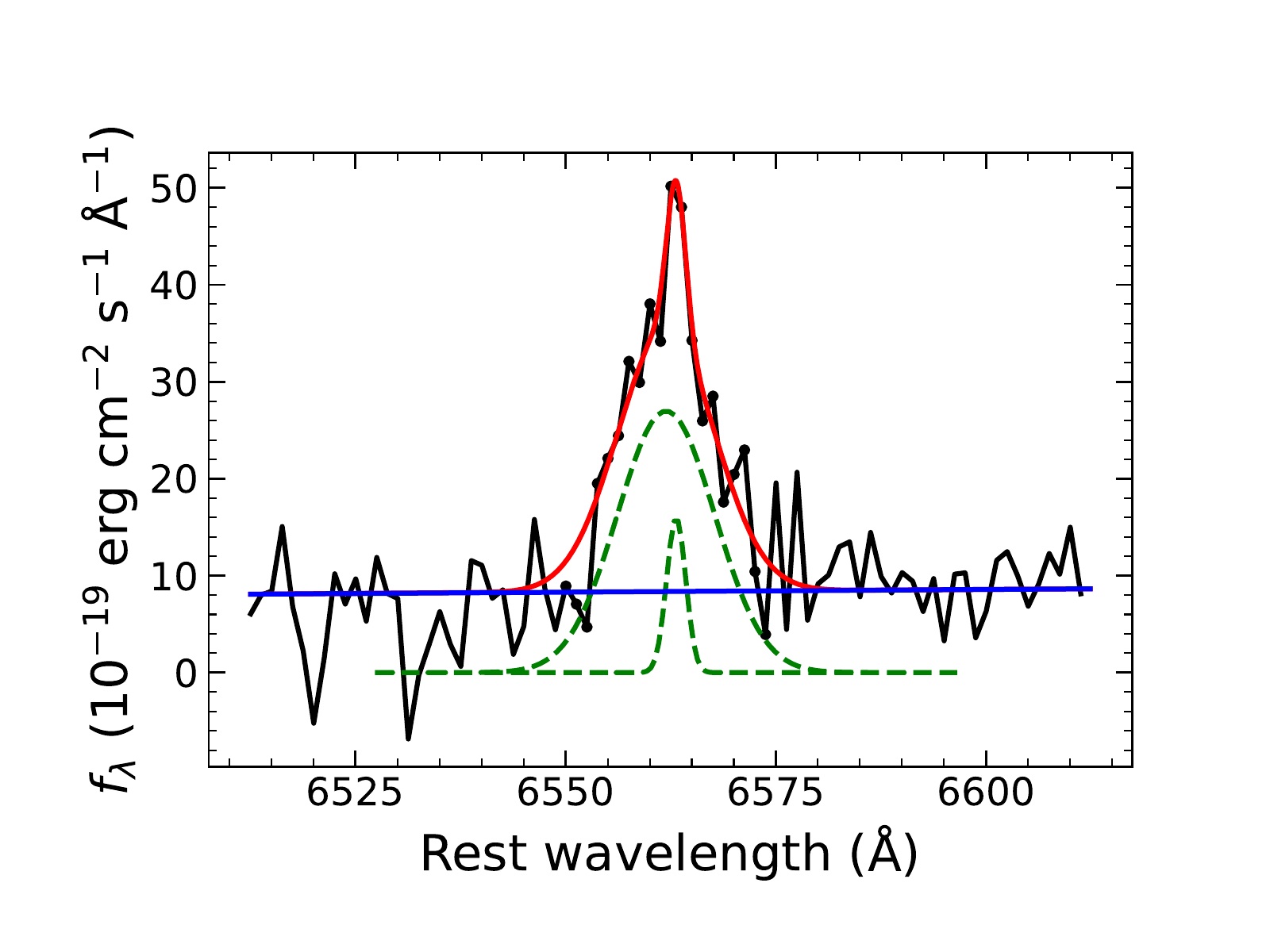}
\caption{{\bf Top}: optical spectrum for the ULX optical counterpart. {\bf Bottom}: spectral decomposition for \halpha. The blue line represents the continuum component. Dashed green curves are the two Gaussians, respectively, for the narrow and broad components. The red curve is the summed model spectrum. Data points with dots are used for the line fit.}
\label{fig:spec_ulx}
\end{figure}

\tabletypesize\footnotesize
\begin{deluxetable}{lccc}
\tablecaption{Emission line properties for the three bubbles.}
\label{tab:line_bubbles}
\tablewidth{\columnwidth}
\tablehead{
\colhead{Line} & \colhead{$f$} & \colhead{$v$} & \colhead{FWHM} \\
\colhead{} & \colhead{($10^{-16}$~\ergcms)} & \colhead{(\kms)} & \colhead{(\kms)} 
}
\startdata
\multicolumn{4}{c}{A}\\
\noalign{\smallskip}\hline\noalign{\smallskip}  
\HeIIwave & $4.2 \pm 0.6$  & $31.0 \pm 12.0$ &  $220 \pm 40$ \\
\hbeta    & $11.5 \pm 0.3$  & $17.0 \pm 1.2$ &  $<41$ \\
\OIIIa    & $10.7 \pm 0.1$  & $18.8 \pm 0.7$ &  $27 \pm 10$ \\
\OIIIb    & $31.9 \pm 0.4$  & $18.8 \pm 0.7$ &  $27 \pm 10$ \\
\NIIa     & $3.1 \pm 0.5$  & $20.0 \pm 6.0$ &  $<78$ \\
\halpha   & $49.8 \pm 0.7$  & $22.6 \pm 0.5$ &  $44 \pm 4$ \\
\NIIb     & $10.4 \pm 0.1$  & $26.5 \pm 0.4$ &  $<20$ \\
\SIIa     & $20.3 \pm 0.3$  & $30.0 \pm 0.5$ &  $37 \pm 4$ \\
\SIIb     & $14.2 \pm 0.2$  & $30.0 \pm 0.5$ &  $37 \pm 4$ \\
\noalign{\smallskip}\hline\noalign{\smallskip}  
\multicolumn{4}{c}{A $-$ C}\\
\noalign{\smallskip}\hline\noalign{\smallskip}  
\HeIIwave  &  $  3.5 \pm 0.5 $    &  $  40.0 \pm 13.0 $ &  $ 210 \pm 40 $    \\
\hbeta     &  $  9.7 \pm 0.3 $    &  $ 18.8 \pm 1.5 $   &        $<45$       \\
\OIIIa     &  $ 8.2  \pm 0.1 $    &  $ 23.8 \pm 0.8 $   &        $<33$       \\
\OIIIb     &  $ 24.5 \pm 0.4 $    &  $ 23.8 \pm 0.8 $   &        $<33$       \\
\NIIa      &  $ 2.5 \pm 0.5  $    &  $ 19.0 \pm 7.0 $   &        $<86$       \\
\halpha    &  $ 40.5 \pm 0.3 $    &  $ 24.1 \pm 0.3 $   &  $  36 \pm 3  $    \\
\NIIb      &  $ 8.8  \pm 0.2 $    &  $ 27.5 \pm 0.7 $   &        $<26$       \\
\SIIa      &  $ 16.2 \pm 0.3 $    &  $ 31.2 \pm 0.7 $   &  $ 25 \pm 7   $    \\
\SIIb      &  $ 11.3 \pm 0.3 $    &  $ 31.2 \pm 0.7 $   &  $  25 \pm 7  $    \\
\noalign{\smallskip}\hline\noalign{\smallskip}  
\multicolumn{4}{c}{B}\\
\noalign{\smallskip}\hline\noalign{\smallskip}  
\HeIIwave  &       $<0.5$         &       \nodata       &    \nodata         \\     
\hbeta     &  $  4.3 \pm 0.3 $    &  $ 16.0 \pm 4.0 $   &  $ 50 \pm 40  $    \\
\OIIIa     &  $ 8.1  \pm 0.1 $    &  $ 29.1 \pm 1.1 $   &  $ 115 \pm 5  $    \\
\OIIIb     &  $ 24.0 \pm 0.4 $    &  $ 29.1 \pm 1.1 $   &  $ 115\pm 5   $    \\
\NIIa      &  $ 0.8 \pm 0.1  $    &  $ 26.3 \pm 2.7 $   &  $ 40 \pm 20  $    \\
\halpha    &  $ 16.2 \pm 0.1 $    &  $ 21.1 \pm 0.4 $   &  $ 103\pm 1   $    \\
\NIIb      &  $ 2.4  \pm 0.1 $    &  $ 26.9 \pm 1.7 $   &  $ 97 \pm 6   $    \\
\SIIa      &  $ 5.0  \pm 0.1 $    &  $ 23.8 \pm 1.0 $   &  $ 99 \pm 4   $    \\
\SIIb      &  $ 3.6  \pm 0.1 $    &  $ 23.8 \pm 1.0 $   &  $  99 \pm 4  $    \\
\noalign{\smallskip}\hline\noalign{\smallskip}  
\multicolumn{4}{c}{C}\\
\noalign{\smallskip}\hline\noalign{\smallskip}  
\HeIIwave  &  $  0.7 \pm 0.2 $    &  $ -27.0 \pm 19.0 $ &  $ 210 \pm 60 $    \\
\hbeta     &  $  1.8 \pm 0.1 $    &  $  8.1 \pm 1.3 $   &      $<45$         \\
\OIIIa     &  $ 2.5  \pm 0.0 $    &  $ -0.3 \pm 0.9 $   &  $ 66  \pm 6  $    \\
\OIIIb     &  $ 7.5  \pm 0.1 $    &  $ -0.3 \pm 0.9 $   &  $ 66 \pm 6   $    \\
\NIIa      &  $ 0.6 \pm 0.0  $    &  $ 26.8 \pm 0.2 $   &  $ 81 \pm 1   $    \\
\halpha    &  $ 9.4  \pm 0.4 $    &  $ 14.1 \pm 1.7 $   &  $ 71 \pm 8   $    \\
\NIIb      &  $ 1.7  \pm 0.1 $    &  $ 18.8 \pm 2.3 $   &  $ 72 \pm 11  $    \\
\SIIa      &  $ 4.1  \pm 0.1 $    &  $ 23.5 \pm 1.2 $   &  $ 73 \pm 5   $    \\
\SIIb      &  $ 2.9  \pm 0.1 $    &  $ 23.5 \pm 1.2 $   &  $  73 \pm 5  $    \\
\enddata
\tablecomments{90\% upper limits are quoted if the parameter cannot be determined.}
\end{deluxetable}

The optical spectra extracted from the bubbles are displayed in Figure~\ref{fig:spec_bubbles}. Significant emission lines are fitted with Gaussians locally in the same manner as described above, with the flux, velocity, and velocity dispersion listed in Table~\ref{tab:line_bubbles}. 

The optical counterpart\footnote{In this work, the optical counterpart refers to the optical object with emission originating from the ULX binary system.} of NGC 55 ULX-1 has been identified by aligning the Chandra and HST images \citep{Gladstone2013}. Based on the error circle, they found two potential optical counterparts in HST images. Their ``source 2'' does not display any detectable emission line and is only visible in the continuum image. ``Source 1'' shows a number of emission lines and in particular a broad \halpha, and we determine it as the optical counterpart to ULX-1. The optical spectrum of the optical counterpart is extracted from a circular region centered on it with a radius of 0\farcs6, and the background is estimated from a nearby region within Bubble A where the \halpha\ surface brightness is similar. The spectrum is shown in Figure~\ref{fig:spec_ulx}, with a decomposition of \halpha. The \halpha\ emission line contains a broad component. We thus fit it with two Gaussians on top of a continuum. The continuum was first determined using data in 6475--6500~\AA\ and 6625--6650~\AA\ and subtracted. The two Gaussians are for the narrow and broad components, respectively. For the narrow component, the intrinsic width cannot be well constrained and we thus fixed it at the typical nebular value of 30~\kms. We tried different background regions where the nebular contribution to the narrow component and continuum may vary, but found that we obtained consistent results for the broad component, which has a FWHM of $600 \pm 60$~\kms\ and an equivalent width of about 50~\AA. We note that the continuum level is rather low, and highly dependent on how one chooses the background region. Thus, the equivalent width may come with large uncertainties up to a factor of 3. 

\section{Discussion}
\label{sec:discuss}

In this work, we present VLT MUSE observations of NGC 55 ULX-1 and its local environment. Three bubble nebulae as well as the ULX optical counterpart are identified. The projected extension/diameter is about 180~pc $\times$ 110~pc for Bubble A, and 80~pc for Bubble B, similar to but slightly smaller than bubbles found around other ULXs \citep{Pakull2002,Pakull2003}. Based on the measured \halpha/\hbeta\ flux ratio and assuming an intrinsic Balmer decrement \halpha/\hbeta\ = 2.95--2.99 \citep{Allen2008}, the total extinction including the Galactic contribution is derived as $E(B-V)$ = 0.38, 0.34, 0.25, and 0.59, respectively, for Bubbles A, A $-$ C, B, and C. We adopt these values in the following analysis and assume a Galactic extinction law \citep{Cardelli1989} for extinction correction.

\tabletypesize\scriptsize
\begin{deluxetable*}{lcccccccccccc}
\tablecaption{Measured and MAPPINGS III simulated emission line flux ratios.}
\label{tab:ratio}
\tablewidth{\textwidth}
\tablehead{
\colhead{Ratio} & 
\colhead{A} & \colhead{300$^\dagger$} & \colhead{275$^\ddagger$} & 
\colhead{A $-$ C} & \colhead{300$^\dagger$} & \colhead{250$^\ddagger$} & 
\colhead{B} & \colhead{425$^\dagger$} & \colhead{350$^\ddagger$} & 
\colhead{C} & \colhead{375$^\dagger$} & \colhead{300$^\ddagger$}
}
\startdata
\OIIIb/\hbeta & $2.78 \pm 0.08$ & $2.64$ & $3.33$ & $2.53 \pm 0.08$ & 2.64 & 2.20 & $5.64 \pm 0.43$ & 5.60 & 5.59 & $4.20 \pm 0.15$ & 4.05 & 4.11 \\
\NIIb/\halpha & $0.21 \pm 0.01$ & $0.83$ & $0.74$ & $0.22 \pm 0.01$ & 0.83 & 0.67 & $0.15 \pm 0.01$ & 0.99 & 0.78 & $0.18 \pm 0.01$ & 0.96 & 0.78 \\
\SII/\halpha & $0.69 \pm 0.01$ & $0.57$ & $0.33$ & $0.68 \pm 0.01$ & 0.57 & 0.29 & $0.53 \pm 0.01$ & 0.61 & 0.37 & $0.75 \pm 0.03$ & 0.62 & 0.35 \\
\enddata
\tablenotetext{\dagger}{Shock velocities in \kms\ that best match the observed \OIIIb/\hbeta\ given $n_0 = 0.1$~\cm.}
\tablenotetext{\ddagger}{Shock velocities in \kms\ that best match the observed \OIIIb/\hbeta\ given $n_0 = 0.01$~\cm.}
\end{deluxetable*}

\begin{figure}
\centering
\includegraphics[width=\columnwidth]{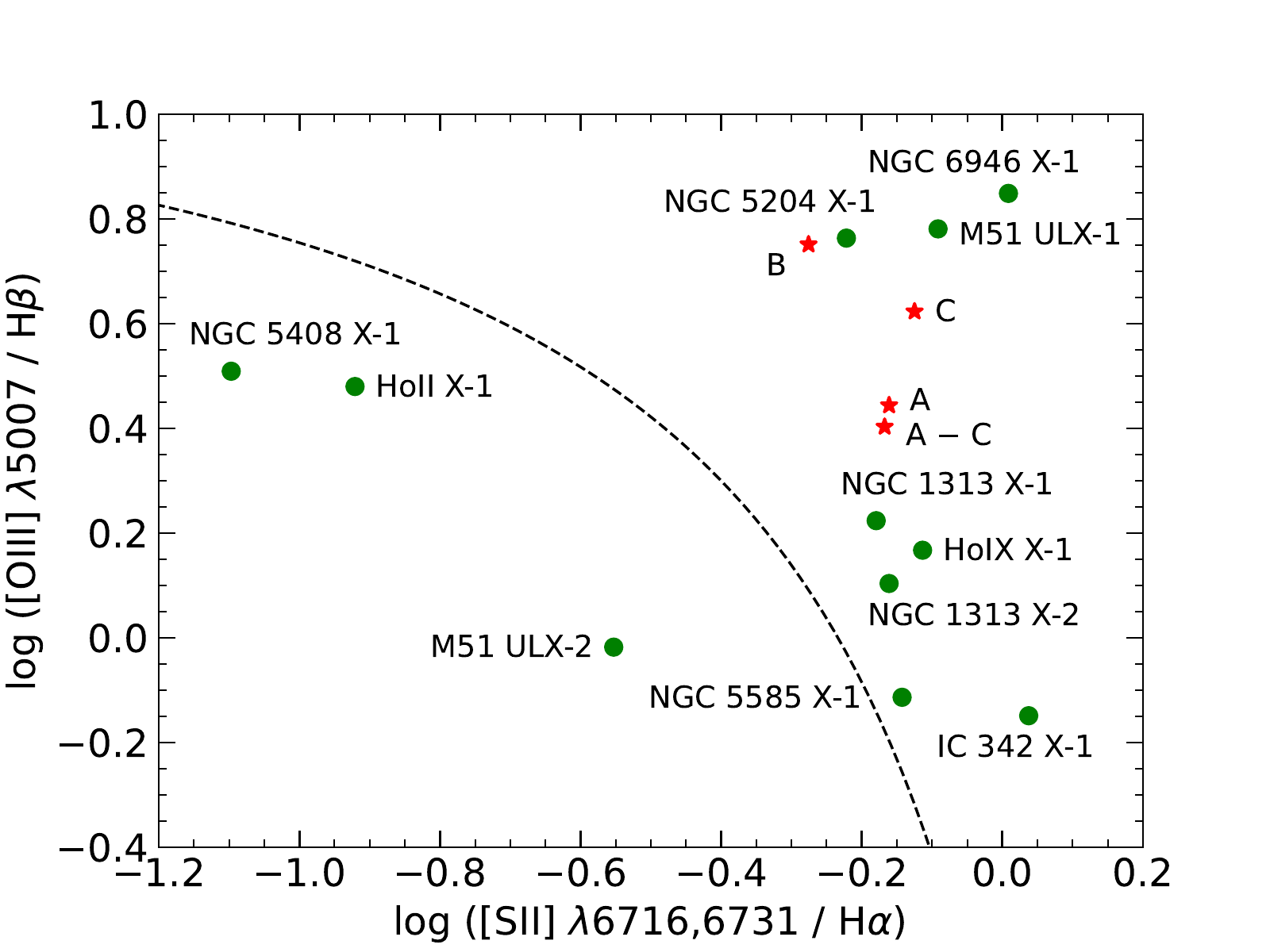}
\caption{The emission line diagnostic diagram \citep{Baldwin1981} for Bubbles A, A $-$ C, B, and C (stars), and other ULX nebulae (circles):
Holmberg II X-1 \citep{Lehmann2005},
Holmberg IX X-1 \citep{Abolmasov2007},
IC 342 X-1 \citep{Roberts2003},
M51 ULX-1 \& ULX-2 \citep{Urquhart2018},
NGC 1313 X-1 \citep{Gurpide2022} and X-2 \citep{Zhou2022},
NGC 5204 X-1 \citep{Abolmasov2007},
NGC 5408 X-1 \citep{Kaaret2009,Cseh2012},
NGC 5585 X-1 \citep{Soria2021},
and NGC 6946 X-1 \citep{Abolmasov2007}.
The dashed curve adopted from \citet{Kewley2001} separates photoionization and shock-ionization.}
\label{fig:bpt}
\end{figure}

The velocity dispersion is typically $\sim$30~\kms\ for Bubble A or A $-$ C, $\sim$100~\kms\ for Bubble B, and $\sim$70~\kms\ for Bubble C. The relatively high velocity dispersions in particular for Bubbles B and C suggest that they are more likely due to shock-ionization than photoionization. Even for Bubble A, which has the smallest velocity dispersion, its regular elliptical shape \citep[similar to the bubble around NGC 1313 X-2;][]{Pakull2002,Zhou2022} indicates that it is powered by winds from the ULX. In Figure~\ref{fig:bpt}, we plot the diagnostic diagram using emission line flux ratios \citep{Baldwin1981} for the bubbles together with other known ULX nebulae. As one can see, the bubbles around NGC 55 ULX-1 belong to the shock-ionization regime, although such an approach is illustrative and may have a large uncertainty. 

The emission line flux ratios for the bubbles are listed in Table~\ref{tab:ratio} and compared with theoretical predictions due to shock-ionization based on MAPPINGS III simulations \citep{Allen2008}. In Table~\ref{tab:ratio}, we quoted the tabulated MAPPINGS III results that best match the observed \OIIIb/\hbeta. Assuming solar abundance, an interstellar medium (ISM) number density $n_0 = 0.1$~\cm, and a magnetic field $B = 1$~$\mu$G, a shock velocity of $\sim$300~\kms, $\sim$300~\kms, $\sim$425~\kms, and $\sim$375~\kms\ can explain the observed \OIIIb/\hbeta\ flux ratio seen in Bubbles A, A $-$ C, B, and C, respectively. In all the cases, the predicted \SII/\halpha\ can also provide a reasonable fit with observations. We note that the same shock velocities and similar line flux ratios can be derived if one assumes $n_0 = 1$~\cm.  The observed \NIIb/\halpha\ is significantly lower compared with simulations, similar to the situation seen in other ULX bubble nebulae, e.g., NGC 1313 X-1 \citep{Gurpide2022} \& X-2 \citep{Ripamonti2011,Zhou2022} and NGC 5585 X-1 \citep{Soria2021}, suggestive of a lower abundance of nitrogen. The nature of nitrogen deficit is unclear. If we assume a subsolar abundance in MAPPINGS III to match the observed \NIIb/\halpha, the \SII/\halpha\ ratio cannot be reconciled. If we assume $n_0 = 0.01$~\cm, the best-fit shock velocities are reduced to $\sim$275~\kms, $\sim$250~\kms, $\sim$350~\kms, and $\sim$300~\kms, respectively, for Bubbles A, A $-$ C, B, and C. The predicted line flux ratios in these cases are also listed in Table~\ref{tab:ratio}. As one can see, with $n_0 = 0.01$~\cm, the \SII/\halpha\ is moderately underestimated, and an overestimation of \NIIb/\halpha\ still exists with a reduced tension. 

For a pressure driven bubble, the bubble radius ($R$), shock velocity ($v_{\rm s}$), and age ($t$) are related as \citep{Weaver1977}
\begin{align}
R &= 0.76 \, t^\frac{3}{5} \left( \frac{P}{\rho_0} \right)^\frac{1}{5} \, ,\\
v_{\rm s} &= 0.38 \left( \frac{P}{R^2 \rho_0} \right)^\frac{1}{3} \, , {\rm and} \\
t &= \frac{3R}{5v_{\rm s}} \, ,
\end{align}
where $P$ is the total jet/wind mechanical power and $\rho_0$ is the pre-shock ISM density. The ISM number density can be inferred from the \hbeta\ surface brightness and the shock velocity as \citep{Dopita1996}
\begin{equation}
\frac{S_{\rm H\beta}}{{\rm erg \, s}^{-1} \, {\rm cm}^{-2}} = 7.44 \times 10^{-6} \left( \frac{{v_{\rm s}}}{10^2\, {\rm km\, s}^{-1}} \right)^{2.41} \left( \frac{n_0}{{\rm cm}^3} \right) \, ,
\label{equ:density}
\end{equation}
where $S_{\rm H\beta}$ can be found from the \hbeta\ luminosity as $S_{\rm H\beta} = L_{\rm H\beta} / (4 \pi R^2)$. Assuming solar abundance with neutral gas, we have the average atomic weight $\mu = 1.2$ \citep{Asplund2009} and $\rho_0 = \mu m_{\rm p} n_0$, where $m_{\rm p}$ is the proton mass.  Plugging in the observed \hbeta\ luminosity, bubble size, and the inferred shock velocity from MAPPINGS III simulations, we obtained the wind power and age in Table~\ref{tab:power}, with two sets of shock velocities derived with $n_0 = 0.1$ and 0.01~\cm, respectively. The density derived from Eq.~(\ref{equ:density}) is relatively low, in the range of 0.01--0.04~\cm, more consistent with the $n_0 = 0.01$~\cm\ assumption. As discussed above, to fit the observed \OIIIb/\hbeta\ and \SII/\halpha\ simultaneously, a relatively high density of 0.1--1~\cm\ is needed. To solve this problem, a dedicated modeling of the nebulae with both shock-ionization and photoionization is perhaps needed. The current data suggest that the true density should be not far away from the range of 0.1--1~\cm.

In Bubbles A (or A $-$ C) and C, the \HeIIwave\ emission is found to be much broader than other lines.  In Figure~\ref{fig:img}, the \HeIIwave\ shows a distribution concentrated toward the ULX position. The image also suggests that the \HeII\ flux measured in Bubble C is in fact ``leaked'' from Bubble A. This suggests that the broad \HeII\ may have reflected a new component of gas in the shock heated low-density bubble interior \citep{Weaver1977,Siwek2017} to a high temperature and ionization state. We note that photoionization usually produces a narrow line width for high ionization species, e.g., as one can see in the nebula around M51 ULX-1 \citep{Urquhart2018}. Thus, the broad \HeII\ emission is not due to photoionization.

\tabletypesize\scriptsize
\begin{deluxetable}{cccccc}
\tablecaption{Physical properties of the three bubbles.}
\label{tab:power}
\tablewidth{\columnwidth}
\tablehead{
\colhead{Bubble} & \colhead{$P$} & \colhead{$R$} & \colhead{$t$} & \colhead{$v_{\rm s}$} & \colhead{$n_0$} \\
\colhead{} & \colhead{($10^{38}$~\ergs)} & \colhead{(pc)} & \colhead{(Myr)} & \colhead{(\kms)} & \colhead{(cm$^{-3}$)} 
}
\startdata
\noalign{\smallskip}
\multicolumn{6}{c}{$v_{\rm s}$ derived assuming $n_0 = 0.1$~\cm}\\
\noalign{\smallskip}\hline\noalign{\smallskip} 
A & $11.2 \pm 0.3$ & $91 \times 56$ & $0.17$  & $300$  & $0.016 \pm 0.001$ \\
A $-$ C & $8.3 \pm 0.3$ & $91 \times 56$ & 0.17 & 300 & $0.012 \pm 0.001$ \\
B       & $3.3 \pm 0.2$ & 39             & 0.05 & 425 & $0.009 \pm 0.001$ \\
C       & $3.9 \pm 0.2$ & 30             & 0.05 & 375 & $0.025 \pm 0.001$ \\
\noalign{\smallskip}\hline\noalign{\smallskip} 
\multicolumn{6}{c}{$v_{\rm s}$ derived assuming $n_0 = 0.01$~\cm}\\
\noalign{\smallskip}\hline\noalign{\smallskip} 
A & $10.5 \pm 0.3$ & $91 \times 56$ & $0.18$  & $275$  & $0.019 \pm 0.001$ \\  
A $-$ C & $7.2 \pm 0.2$ & $91 \times 56$ & 0.20 & 250 & $0.018 \pm 0.001$ \\
B       & $2.9 \pm 0.2$ & 39             & 0.06 & 350 & $0.014 \pm 0.001$ \\
C       & $3.4 \pm 0.2$ & 30             & 0.06 & 300 & $0.042 \pm 0.002$ \\
\enddata
\end{deluxetable}

Bubble A is in good analogy to the bubble nebula surrounding NGC 1313 X-2 \citep{Pakull2002, Pakull2003,Zampieri2004,Mucciarelli2005,Pakull2006,Ramsey2006,Ripamonti2011,Zhou2022}. The mechanical power needed to power it is close to the X-ray luminosity, both on the order of $10^{39}$~\ergs. This nebula or all the surrounding nebulae seem to have the least mechanical power among ULX bubbles, consistent with the fact that its X-ray luminosity is just above the ULX threshold. The closest case would be NGC 51 ULX-1, which has a mechanical power and X-ray luminosity both around $2 \times 10^{39}$~\ergs\ \citep{Urquhart2016,Urquhart2018}. However, as the observed X-ray flux may be beamed toward our line of sight or heavily obscured by the thick disk or wind, we do not expect a tight correlation between them, e.g., the X-ray luminosities of SS~433 and NGC 7793 S26 are far less than the mechanical powers of their surrounding nebulae.

Bubble B projects 11\arcsec\ (or at least 95~pc) to the ULX on the sky plane, and is apparently not contiguous to Bubble A.  Its morphology is reminiscent of a bow shock, with emission enhanced toward the edge away from the ULX direction, see the \halpha\ and \OIII\ images in Figure~\ref{fig:img}. Therefore, we argue that Bubble B is plausibly inflated by a collimated jet launched from the ULX, similar to the nebula around NGC 7793 S26 \citep{Pakull2010,Soria2010}, Cygnus X-1 \citep{Gallo2005}, SS~433/W50 \citep{Fabrika2004}, or radio lobes from active galactic nuclei. 

If this is the case, we predict that there should be detectable radio emission around Bubble B. Assuming synchrotron emission under the minimum-energy approximation, the radio specific flux is roughly scaled with the jet power, age, and medium density as \citep{Soria2010}
\begin{equation}
S_\nu \propto P^{1.34} t^{0.32} n_0^{0.51} \, .
\end{equation}
Taking $S_\nu = 2.1$~mJy, $P = 5 \times 10^{40}$~\ergs, $t = 0.2$~Myr, $n_0 = 0.7$~\cm, and a distance of 3.9~Mpc for NGC 7793 S26 \citep{Pakull2010,Soria2010}, we estimate a total radio flux of roughly 1--3~$\mu$Jy at 5~GHz for Bubble B. We may also make a direct comparison with the radio nebula W50 around SS~433, where the jets have a similar power of $10^{39}$~\ergs\ as in our case. W50 has a flux density of 34~Jy at 4.75~GHz \citep{Downes1986}, and would appear to be about 200-300 $\mu$Jy if it is located at a distance of 1.78~Mpc instead of 5~kpc \citep{Fabrika2004}. Therefore, the radio emission in Bubble B is estimated to have a specific flux density of about $1 - 10^2$~$\mu$Jy.


The nature of Bubble C is unclear. If Bubble B is inflated by a collimated jet, it would be difficult to explain Bubble C as a result of another jet, and vice versa. There is no peculiarity around the Bubble C region in HST broad band images. We cannot rule out that Bubble C is simply a part of Bubble A. The broad line width seen in it could be a consequence of low density in this region. 

A transient X-ray source T2 with a peak luminosity of $2 \times 10^{38}$~\ergs\ appears on the sky plane near the ULX \citep{Jithesh2016}. The angular separation is about 90\arcsec, corresponding to a projected distance of 780 pc, from T2 to the bubbles. This is much longer than any known jet powered nebulae such as those around Cygnus X-1 \citep[$\sim$5 pc;][]{Gallo2005}, SS 433 \citep[$\sim$100 pc;][]{Fabrika2004}, and NGC 7793 S26 \citep[$\sim$150 pc;][]{Pakull2010}.  Thus, it is unlikely that there is a physical connection between T2 and the bubbles.

The \halpha\ emission in the spectrum of the ULX optical counterpart contains a broad component with a FWHM of about 600~\kms\ and an equivalent width of roughly 50\AA, similar to the \halpha\ line seen in many other ULXs \citep{Fabrika2015}. The \HeIIwave\ line is barely seen in the spectrum; the low S/N does not allow us to characterize its properties. One possible origin of the emission lines could be the hot wind as suggested by \citet{Fabrika2015}. In that case, \HeII\ arises from a location closer to the central compact object in an accelerating wind and should have a velocity dispersion smaller than \halpha. We examined the spectrum and found that a \HeII\ line width of $< 600$~\kms\ is consistent with the low S/N data. 

We note that the above analysis and discussions mainly count on the emission line intensity and flux ratios. Useful information encoded in the velocity dispersion and morphology has not been fully utilized. For a complete understanding of these bubble nebulae, future 3D numerical simulations are needed. 

\begin{acknowledgments}
We thank the anonymous referee for useful comments. HF acknowledges funding support from the National Key R\&D Project under grant 2018YFA0404502, the National Natural Science Foundation of China under grants Nos.\ 12025301, 12103027, \& 11821303, and the Tsinghua University Initiative Scientific Research Program.
\end{acknowledgments}


\end{document}